\documentclass[floatfix,aps,prl,twocolumn,superscriptaddress,nofootinbib]{revtex4-1}

\usepackage[draft]{todonotes}
\usepackage{csquotes}
\usepackage{amsmath}
\usepackage{color}
\usepackage{dcolumn}

\newcommand\UCLA{Dept. of Physics and Astronomy, Univ. of California, Los Angeles, Los Angeles CA 90095, USA}
\newcommand\JPL{Jet Propulsion Laboratory, California Institute of Technology, Pasadena CA 91109, USA}
\newcommand\UD{Dept. of Physics, Univ. of Delaware, Newark DE 19716, USA.}

\newcommand\CalPoly{Dept. of Physics, California Polytechnic State Univ., San Luis Obispo CA 93407, USA}
\newcommand\KIT{Institut f\"{u}r Astroteilchenphysik, Karlsruher Institut f\"{u}r Technologie, 76344 Eggenstein-Leopoldshafen, Germany}
\newcommand\KITEKP{Institut f\"{u}r Experimentelle Teilchenphysik, Karlsruher Institut f\"{u}r Technologie, 76128 Karlsruhe, Germany}
\newcommand\Kavli{Kavli Institute for Cosmological Physics, Univ. of Chicago, Chicago IL 60637, USA}
\newcommand\NTU{Dept. of Physics, Grad. Inst. of Astrophys., National Taiwan University, Taipei Taiwan}
\newcommand\UH{Dept. of Physics and Astronomy, Univ. of Hawaii, Manoa HI 96822, USA.}
\newcommand\SLAC{SLAC National Accelerator Laboratory, Menlo Park CA, 94025, USA}

\newcommand\WM{Physics Dept., College of William \& Mary, Williamsburg VA 23187, USA.}
\newcommand\UCL{Dept. of Physics. Univ. College London, United Kingdom}
\newcommand\WashU{Dept. of Physics, Washington Univ. in St. Louis, St. Louis MO 63130, USA}
\newcommand\Fermi{Dept. of Physics, Enrico Fermi Institute, Univ. of Chicago, Chicago IL 60637, USA}
\newcommand\Stanford{Dept. of Physics, Stanford Univ., Stanford CA, 94305, USA}
\newcommand\IIHE{Astrophysical Institute, Vrije Universiteit Brussel, Pleinlaan 2, 1050 Brussels, Belgium}
\newcommand\IAP{Sorbonne Universit\'{e}, CNRS, UMR 7095, Institut d’Astrophysique de Paris, 75014 Paris, France}
\newcommand\Penn{Dept. of Physics and Astronomy and Astrophysics, Pennsylvania State University, University Park PA 16802 USA}
\newcommand\RU{Department of Astrophysics/IMAPP, Radboud University, PO Box 9010, 6500 GL, The Netherlands}
\newcommand\RUHEP{Department of High Energy Physics/IMAPP, Radboud University, PO Box 9010, 6500 GL, The Netherlands}
\newcommand\Nikhef{Nikhef, Science Park Amsterdam, Amsterdam, The Netherlands}
\newcommand\NYMCT{Department of Electrophysics, National Yang Ming Chiao Tung University, Hsinchu, Taiwan}

\begin{document}

\title{The SLAC T-510 experiment for radio emission from particle showers: detailed simulation study and interpretation}




\author{K.~Bechtol}\affiliation{\Kavli}
\author{K.~Belov}\affiliation{\UCLA}\affiliation{\JPL}
\author{K.~Borch}\affiliation{\UCLA}
\author{P.~Chen}\affiliation{\NTU}
\author{J.~Clem}\affiliation{\UD}
\author{P.~Gorham}\affiliation{\UH}
\author{C.~Hast}\affiliation{\SLAC}
\author{T.~Huege}\affiliation{\KIT}\affiliation{\IIHE}
\author{R.~Hyneman}\affiliation{\WM}
\author{K.~Jobe}\affiliation{\SLAC}
\author{K.~Kuwatani}\affiliation{\UCLA}
\author{J.~Lam}\affiliation{\UCLA}
\author{T.C.~Liu}\affiliation{\NYMCT}
\author{K.~Mulrey}\thanks{Corresponding author}\email{katharine.mulrey@ru.nl}\affiliation{\IIHE}\affiliation{\RU}\affiliation{\Nikhef}

\author{J.~Nam}\affiliation{\NTU}
\author{C.~Naudet}\affiliation{\JPL}
\author{R.J.~Nichol}\affiliation{\UCL}
\author{C. Paciaroni}\affiliation{\CalPoly}
\author{B.F.~Rauch}\affiliation{\WashU}
\author{A.~Romero-Wolf}\affiliation{\JPL}
\author{B.~Rotter}\affiliation{\UH}
\author{D.~Saltzberg}\affiliation{\UCLA}
\author{H.~Schoorlemmer}\affiliation{\UH}\affiliation{\RUHEP}\affiliation{\Nikhef}
\author{D.~Seckel}\affiliation{\UD}
\author{B.~Strutt}\affiliation{\UCL}
\author{A.~Vieregg}\affiliation{\Kavli}\affiliation{\Fermi}
\author{C.~Williams}\affiliation{\Stanford}
\author{S.~Wissel}\affiliation{\CalPoly}\affiliation{\Penn}
\author{A.~Zilles}\affiliation{\KITEKP}\affiliation{\IAP}


\date{\today}

\begin{abstract}
Over the last several decades, radio detection of air showers has been widely used to detect ultra-high-energy cosmic rays.  We developed an experiment under controlled laboratory conditions at SLAC with which we measured the radio-frequency radiation from a charged particle shower produced by bunches of electrons as primaries with known energy.  The shower took place in a target made of High Density Polyethylene located in a strong magnetic field.  The experiment was designed so that Askaryan and magnetically-induced components of the radio emission could be measured independently.  At the same time, we performed a detailed simulation of this experiment to predict the radio signal using two microscopic formalisms, endpoint and ZHS.  In this paper, we present the simulation scheme and make a comparison with data characteristics such as linearity with magnetic field and amplitude.  The simulations agree with the measurements within uncertainties and present a good description of the data.  In particular, reflections within the target that accounted for the largest systematic uncertainties are addressed.  The prediction of the amplitude of Askaryan emission agrees with measurements to within 5\% for the endpoint formalism and 11\% for the ZHS formalism.  The amplitudes of magnetically-induced emission agree to within 5\% for the endpoint formalism and less than 1\% for the ZHS formalism.  The agreement of the absolute scale of emission gives confidence in state-of-the-art air shower simulations which are based on the applied formalisms.
\end{abstract}

\maketitle

When highly energetic cosmic rays impinge on the Earth's atmosphere, they create extensive air showers consisting of cascades of secondary particles. During the shower development, the shower particles emit a radio signal that can be interpreted as a superposition of charge-excess radiation due to the Askaryan effect~\cite{Askaryan1962, Askaryan1965} and magnetically-induced transverse current radiation, called the geomagnetic effect~\cite{geomagnetic,Scholten:2007ky}.
\newline \indent The interpretation of radio measurements from air showers is based on the comparison of the measured radio signal with detailed simulations of the radio emission. This is a common way to interpret air shower data, but it relies on a complete understanding of the radio emission from particle cascades as well as the ability to model the underlying physics in simulations.  In the last years, the analysis of data measured by radio arrays by comparing simulated and measured air-shower events has shown that simulations can reproduce the extremely detailed radio signal from an air shower (see the reviews~\cite{Huege:2016veh} and~\cite{Schroder:2016hrv}).  The state-of-the-art air shower radio emission simulation codes, CoREAS~\cite{CoREAS} and ZHAireS~\cite{ZHAireS}, are both based on microscopic approaches, the endpoint~\cite{James2011} and the ZHS formalisms~\cite{PhysRevD.45.362,Alvarez-Muniz2010}, respectively.  They calculate the radio emission from air showers on the basis of full Monte-Carlo simulations.  The radiation energies (i.e., total radiated energy by an air shower deposited on the ground) as predicted by ZHAireS and CoREAS in the $30-80$~MHz band have been shown to agree to within 5.2\%~\cite{GOTTOWIK201887}.  Nevertheless, it remained to be demonstrated in a laboratory setting that microscopic simulations are able to predict the absolute scaling and features of radio emission from air showers.  This question needed to be answered to prepare for future high precision experiments in the field of radio detection of air showers.

Measurements made in a laboratory are affected by different systematic uncertainties than air shower experiments and provide a comparison between data and simulations independent of hadronic interaction models, unknown mass-composition of the primary particles or unknown geometry.  With dedicated particle-beam experiments at electron accelerators the study of the radio emission from well defined and pure electromagnetic showers with known primary particle type and primary energy is possible. In addition, a direct comparison of data to simulations using the ZHS and the endpoint formalisms for the same shower can be performed.  This comparison is independent of the underlying air shower simulation programs, CORSIKA and AIRES, which differ in the handling of the refractive index model and thinning algorithms used for the electromagnetic shower.  

To this end, the SLAC T~-~510 experiment was carried out, in which a pure electron beam of known parameters was shot into a dense target, positioned in a variable magnetic field of up to 970 Gauss. The geometry of the experiment was designed so that the resulting shower produced magnetically-induced and Askaryan radiation that could be measured in separate antenna channels. A first validation of the experimental results was presented in~\cite{Belov2016}.  

An integral part of the experiment was the detailed simulation study which included both ZHS and endpoint formalisms in the microscopic calculation of radio emission from particle showers.  In this work we discuss the preparation and the execution of this simulation study in detail as well as the comparison of the simulated results to measured data.  In particular, reflections within the target, which contributed to the amplitude of the measured signal, are addressed for the first time.  In our previous work, we estimated that the added signal strength due to the unknown reflection coefficient at the bottom of the target contributed a 40\% systematic uncertainty to the  measurement, and we found that the measured voltage was systematically larger that the simulated voltage by roughly 35\%.  In this work, we measure the reflection coefficient and account for reflections in the analysis, which brings the agreement of the amplitude between simulations and data to within 11\% for Askaryan emission and 5\% for magnetically-induced emission.

This paper is structured as follows.  First, a description is given of the microscopic modeling processes used in the simulations and details of the experimental set up are provided.  Following this, the modeling of radio emission is discussed, including the impact of inconsistencies in the Geant4 treatment of multiple scattering, the inclusion of a realistic magnetic field, the use of the endpoint and ZHS formalisms, the treatment of the boundary layer of the target, and the resulting Cherenkov-like effects. We then present the application of the simulation set-up to the SLAC T-510 experiment.  This includes comparing the performance of the two formalisms and handling transition radiation.  Finally, we compare the simulations to data.  We address the effects of internal reflections within the target, which were the largest source of uncertainty.  Comparisons between data and simulations are made across the Cherenkov cone, and with different magnetic field strengths.

\section{Experimental set-up}
\begin{figure}
\centering
{\includegraphics[width=0.48\textwidth]{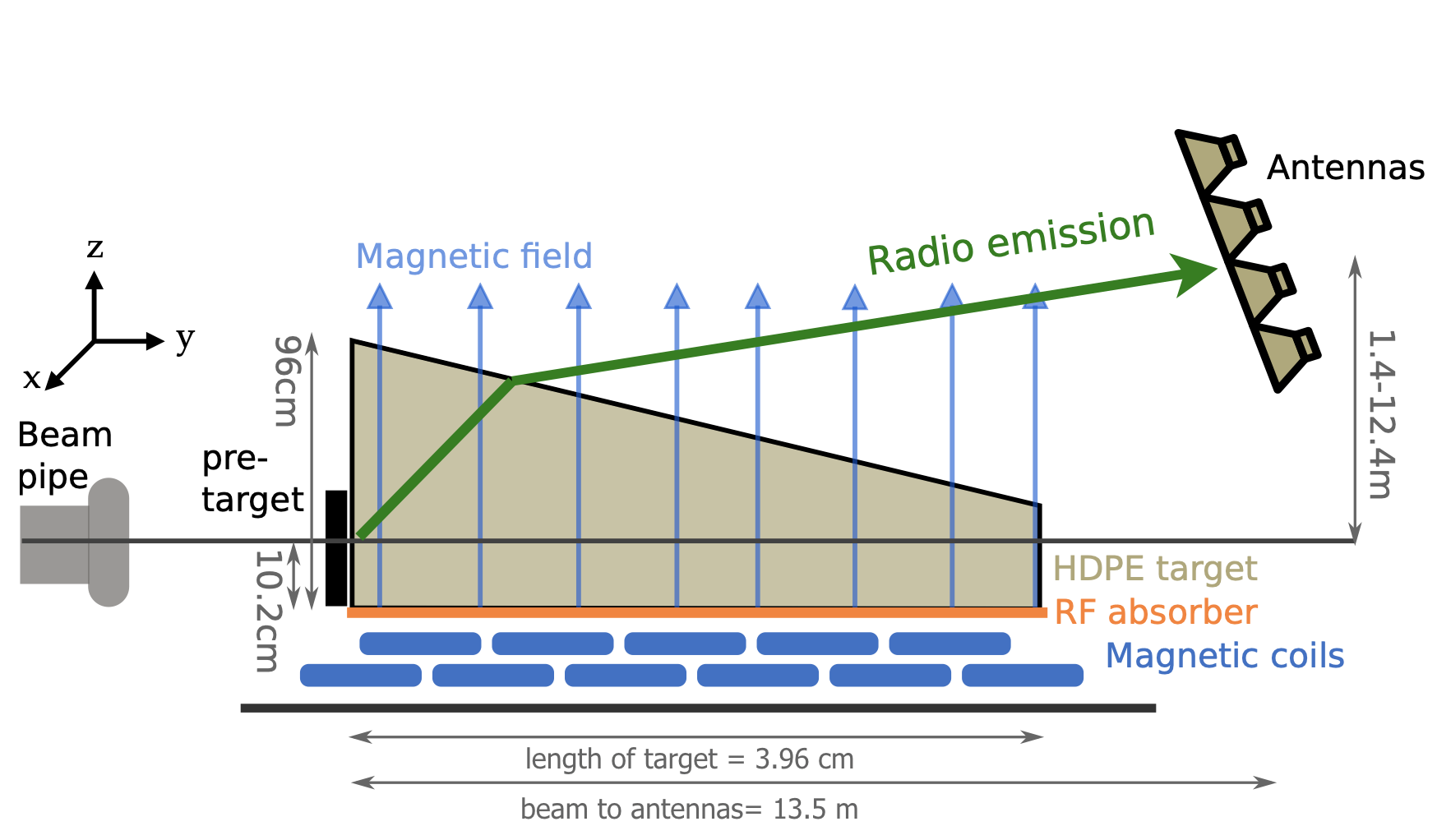}}
\caption{Schematic view of the experimental set-up including the antenna tower position and the geometry for the signal propagation, not to scale.}\label{fig:T510}
\end{figure} 
In January and February of 2014, we performed the T~-~510 experiment at the End Station Test Beam (ESTB) in End Station A (ESA) at the SLAC National Accelerator Laboratory. First results are published in~\cite{Belov2016}.  A schematic of the geometry of the SLAC T-510 experimental set-up is shown in Fig. \ref{fig:T510}.  The SLAC electron beam was shot into a HDPE target with electron energies of $4.35$ and $4.55\,\mbox{GeV}$.  The particle showers generated in the target were equivalent to a shower induced by a primary cosmic ray with an energy of about $4\times 10^{18}\,\mbox{eV}$~\cite{Belov2016}.

To measure the total charge in each bunch shot into the target, an integrated current transformer (ICT) was placed between the end of the beam pipe and the target. Its accuracy is given to within $3\%$ for bunch charges larger than $100\,\mbox{pC}$.   The measured charge distribution had a mean of $131\,\mbox{pC}$ and a standard deviation of $3.3\,\mbox{pC}$ and 2\% systematic uncertainty~\cite{StephICRCProc}.
In addition, a high-frequency S-band ($2-4\,\mbox{GHz}$) horn antenna measured the transition radiation as the beam exited the beam pipe.  This provided a global trigger for the measurement system as well as the shot-to-shot relative calibration of the beam charge.

To reduce the size of the particle shower to laboratory scales, we used a target made out of High-Density Polyethylene (HDPE).  The target was $3.96\,$m long, $0.96\,$m tall, and $0.60\,$m  wide, large enough to contain the vast majority  of the particles in the shower.  The target was set up using $1500$~kg of single bricks, each of a size of $5.08\,\mbox{cm}\times10.16\,\mbox{cm}\times30.48\,\mbox{cm}$. Furthermore, before entering the target, the electron beam passed through a $1.27\,\mbox{cm}$ thick lead plate, acting as a pre-shower medium.

To reduce internal reflections, we positioned the target on an RF absorbing blanket.  In addition, several pieces of RF absorber foam were placed at both sides of the target and at the exit surface of the target, so that the measurement only accounts for radiation exiting trough the upper surface of the target. After analyzing the data we found that the blanket did not prevent reflections from the bottom of the target in our frequency range.  The treatment of these reflections is discussed in this paper.  To avoid total internal reflections of the signal at the top of the target, this surface was chosen to be slanted by an angle of $10.16^\circ$ to the horizontal. The index of refraction of $n_{\text{HDPE}}=1.53$ corresponds to a Cherenkov angle inside the target of $49.2^\circ$. This leads to an expected position of the Cherenkov cone on the vertical axis at about $6.5\,$m above the beam line at a horizontal distance of $13.5\,\mbox{m}$ from the entry point of the beam.

\begin{figure}
\centering
\includegraphics[width=0.47\textwidth]{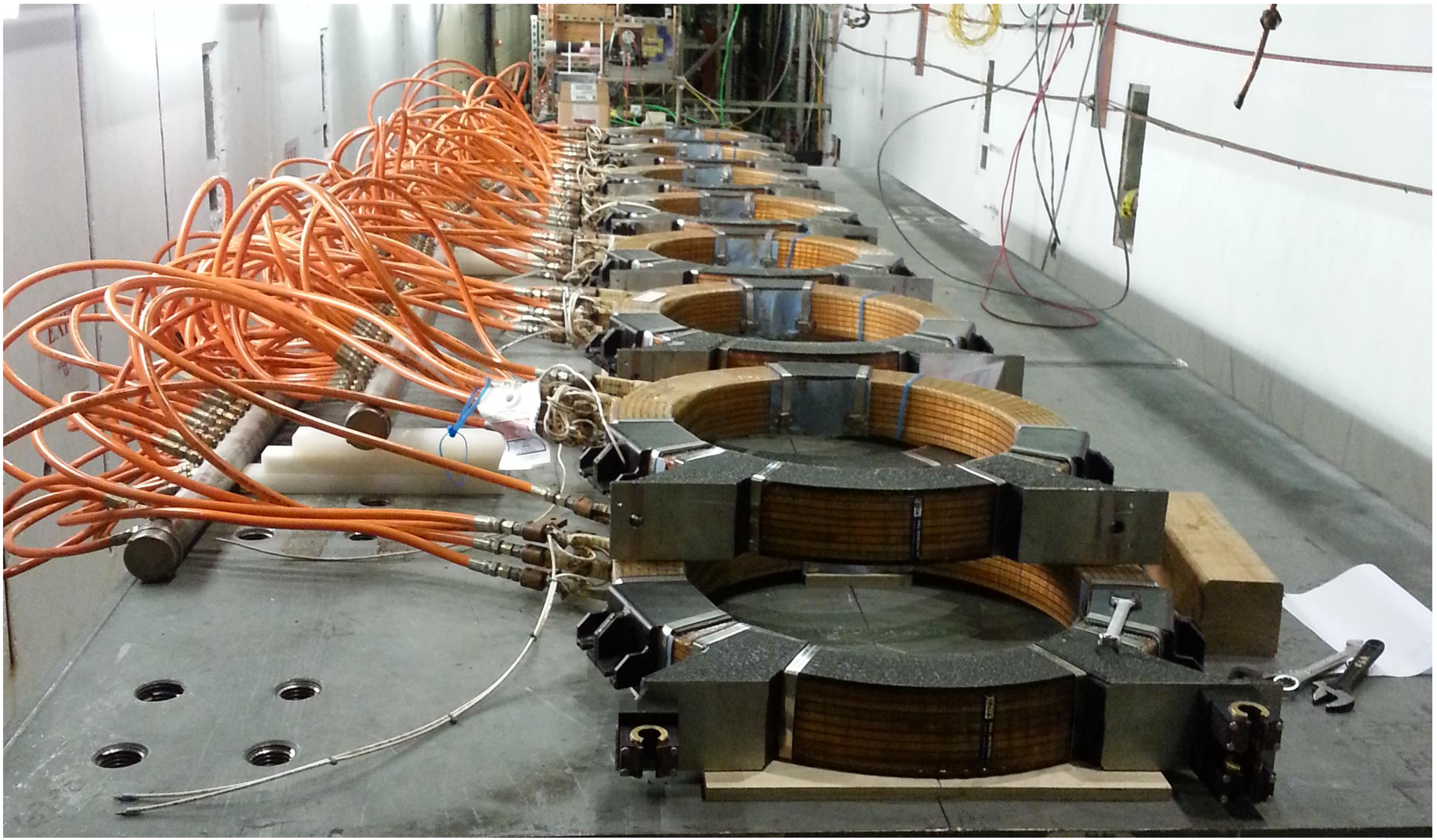}\\
\includegraphics[width=0.48\textwidth]{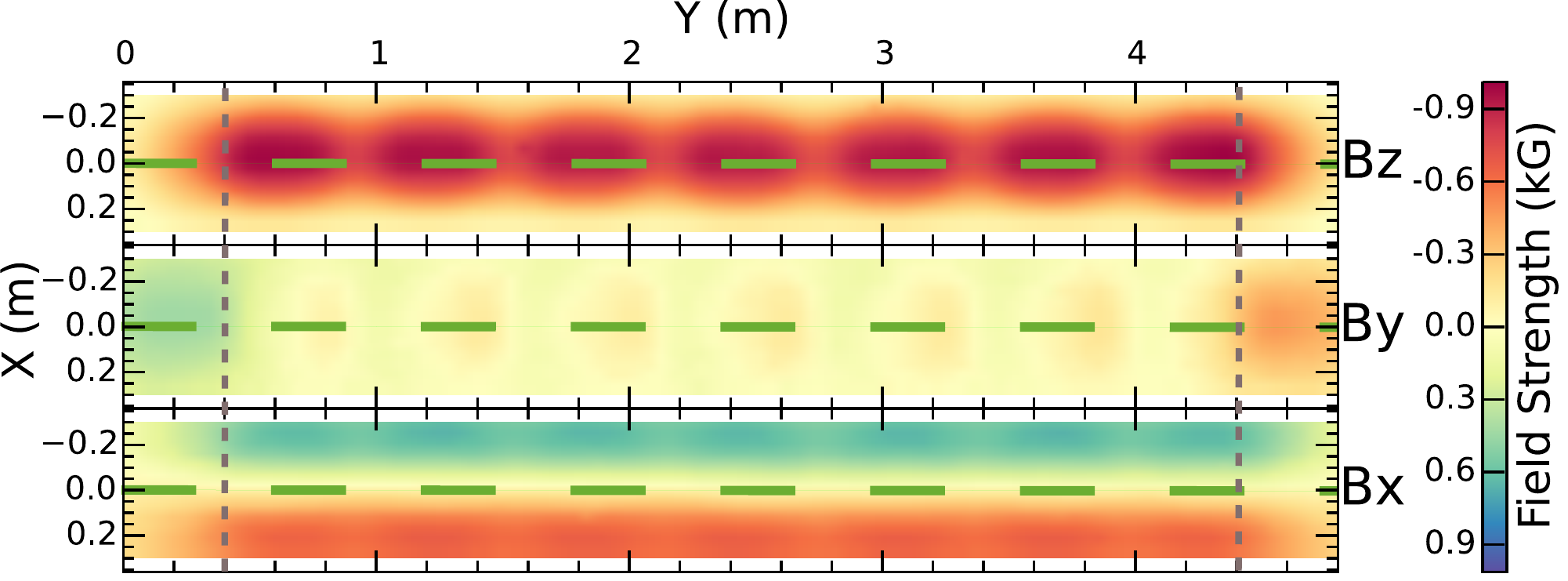}\\
\vspace{0.5cm}
\includegraphics[width=0.4\textwidth]{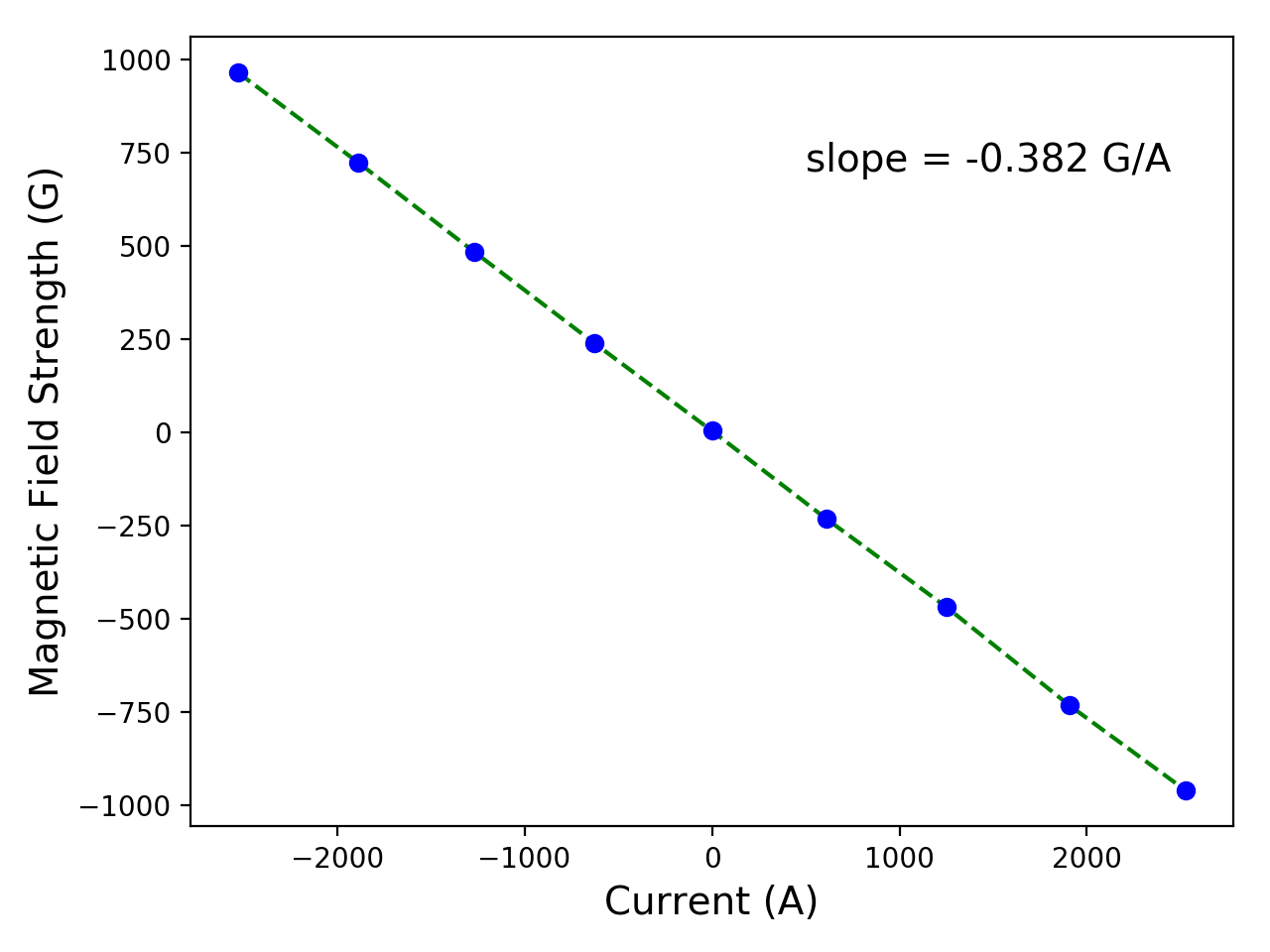}
\caption{Top: Picture of the staggered coils which were used to produce a strong and uniform magnetic field in the vertical direction. Middle: Measured three-dimensional magnetic field map for a current of $2400\,\mbox{A}$ (in a $5\,$cm$\times 5\,$cm grid) which is included in simulations. The dashed gray lines mark the target area, the green lines the position of the beam~\cite{Belov2016} (modified).  Bottom: Measurement of the linear dependence of the magnetic field strength on the applied current.}
\label{fig:Bmap}
\end{figure}

To provide a uniform magnetic field in the vertical direction with a field strength up to $970\,$G, fifteen water-cooled solenoids were staggered in two rows under the target and were supplied with a current of up to $2400\,\mbox{A}$  with reversible polarity during data taking~\cite{Belov2016}.  A picture of the coils and a map of the magnetic field strength distribution are shown in Fig.~\ref{fig:Bmap} (top, middle).  Because the vertical magnetic field component falls off near the edges of the coils, we placed the target as indicated by the dashed lines in this figure.  The strength of the magnetic field along the beam was chosen to be strong enough to bring the expected radiation intensities from the magnetic effect and the Askaryan effect to the same order of magnitude~\cite{Belov2016}.
%

To check for the linearity of the magnetic field strength with the applied current, we measured the magnetic field at beam height at several different currents between $-2400\,\mbox{A}$ and $2400\,\mbox{A}$, shown in Fig.~\ref{fig:Bmap} (bottom). A linear fit to the data returns a scaling factor $m$ for rescaling the complete magnetic field map to the desired strength in dependence on the applied current.
This results in the following linear dependency shown in Fig.~\ref{fig:Bmap} (bottom) with a slope of $m = - 0.382\,\mbox{G/A}$, which we used to set the magnetic field strength within the simulation by specifying the current induced in the coil.

To measure the electric field produced by the particle shower, four dual-polarization, quad-ridged horn antennas~\cite{Gorham:2008dv}, each with an opening of $1\,\mbox{m} \times 1\,\mbox{m}$, were arranged on a frame attached to a crane.  We will refer to the two polarizations of the antennas as vertical and horizontal channels.  The antenna tower was placed at the far wall of ESA in a maximum distance of $L= 13.5\,$m from the entry point of the beam in the target to fulfill the condition for full coherence of the radio emission according to $kL\gg1$,  with  the  wavenumber $k=2\pi n f/c$, the frequency $f$ and the index of refraction $n$~\cite{Lehtinien}.
The antennas were sensitive in a frequency band from $200-1200\,\mbox{MHz}$. The induced signals were digitized with a sampling rate of $5\,\mbox{GSPS}$.

\begin{figure}
\centering
{\includegraphics[width=0.4\textwidth]{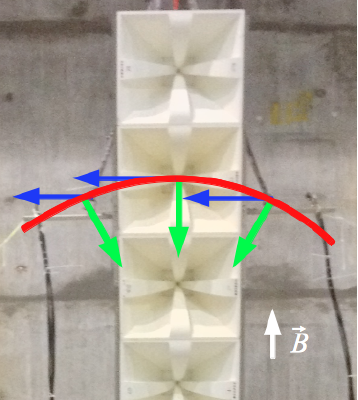}}
\caption{Positioning of the antenna tower at the Cherenkov ring (red) so that contributions of the charge-excess (green) and magnetic effect (blue) are separated into the horizontal and vertical channels of the antenna.}\label{fig:TowerComponents}
\end{figure}

A critical aspect of the T-510 experiment was measuring the magnetically-induced and Askaryan radiation separately. To achieve this, the geometry of the experiment was designed such that the radiation could be separated into horizontally and vertically polarized channels.  The antenna tower was placed on the vertical axis which is perpendicular to the beam axis and parallel to the magnetic field direction (seen Fig.~\ref{fig:T510}).  The Askaryan radiation,  which is due to the time-variation of the net current, is polarized radially. The antenna is aligned so that this radiation is measured in only the vertical channel. The magnetic field induces radiation in the $\vec{v}\times\vec{B}$ direction, where $\vec{v}$ denotes the velocity vector of the shower and $\vec{B}$ denotes the magnetic field.  As shown in Fig.~\ref{fig:Bmap}, the magnetic field was designed to be strongest in the vertical direction.  This orientation was chosen so that the magnetically-induced radiation would be primarily in the horizontal direction.  Thus, the magnetic radiation was observed at the antennas as being horizontally polarized, while the Askaryan radiation was vertically polarized, as shown in Fig.~\ref{fig:TowerComponents}.

%

\section{Microscopic modeling}
In microscopic approaches to the calculation of the radio emission from a particle shower, each single electron and positron is considered separately. In the T-510 experiment, the emitted radio signal is calculated and propagated to the observer position using the endpoint and the ZHS formalisms in parallel. Here, the signals of all particles are superimposed to find the total radio emission from the shower.  Coherence effects and time delays of the emission from individual particles are automatically taken into account with proper handling of the propagation.  This classical electrodynamics calculation of the radio emission does not make assumptions about the emission mechanisms and has no free parameters that would influence the resulting electric field.  In the same way, the endpoint formalism is applied in CORSIKA~\cite{Heck:1998vt}, within the CoREAS extension~\cite{CoREAS}, and the ZHS formalism is built into the AIRES code~\cite{AIRES}, leading to  ZHAireS~\cite{ZHAireS}.  It has been shown that the two approaches are mathematically equivalent~\cite{Radio1,Belov:2013hja}, but numerical aspects can be significantly different.  

\section{Modeling of the radio emission}
Geant4 10.0 was used to simulate the particle shower in the target~\cite{GEANT}. This toolkit is object-oriented and programmed in  C++.  It simulates the passage of particles through matter, handling their propagation and interactions. It treats the shower development by splitting up continuous trajectories of particles into sub-tracks. The sub-tracks can be seen as straight lines with given starting and stopping points and their corresponding times. This information is used as the basis for the calculation of radio emission by particle showers.  The simulation also includes details of the experimental set-up, such as the target geometry and material, as well as the beam energy.  The measured, three-dimensional magnetic field map, as shown in Fig.~\ref{fig:Bmap} (middle) was also included.  All relevant interactions of shower photons, electrons and positrons are properly taken into account.  

For the calculation of the radio emission the charge of a particle, the positions of the sub-track's start- and end-points  and their corresponding times are needed. On this basis, the velocity and acceleration along the sub-track as well as the particle propagation direction can be calculated. The positions of the antennas during the measurements are given as observers for the calculation of the radio signal.

\subsection{Implementation of the realistic magnetic field strength distribution}
To study the effect of the realistic magnetic field on the emission of the radio signal, we included the measured three-dimensional map of the magnetic field strength (shown in Fig.~\ref{fig:Bmap}, middle) as a 3D vector at beam height in the simulation.

Each component of the measured magnetic field map is read in by Geant4. The value of magnetic field strength which affects a particle track is set depending on the position of the current sub-track. Since the magnetic field strength scales linearly with the applied current, the strength of the field in the simulation can be controlled by this dependency on the current set during the measurements (see Fig.~\ref{fig:Bmap}, bottom). The maximum magnetic field strength during the measurements of about $-970\,\mbox{G}$ along the vertical axis along the beam line is given by the maximum applied current of $2400\,\mbox{A}$.

\subsection{Implementation of the emission formalisms in the shower simulation}
The simulations include the calculation of the radio signals produced by the particle showers in the target based on the sub-track positions ($\vec{x}_{\tiny{\mbox{start}}}$,~$\vec{x}_{\tiny{\mbox{end}}}$) and times ($t_{\tiny{\mbox{start}}}$,~$t_{\tiny{\mbox{end}}}$) as given by Geant4. Each sub-track contributes to the calculation of the electric field or to the vector potential using the endpoint and ZHS formalisms which run in parallel. This provides a one-to-one comparison so that shower-to-shower fluctuations are not an issue in the comparison of the results for the two formalisms.
In the simulation, we chose a $400\,\mbox{ns}$ time window for the arrival of the signal at the antenna, starting with the time at which a signal originating from the entry point of the beam to the target would reach the antenna. The sampling rate for the simulated time traces is set to a value of $100\,\mbox{GSPS}$.
The shower simulation is done by injecting $5000\,\mbox{electron primaries}$ with an energy of $4.35$ or $4.55\,\mbox{GeV}$ each. Due to coherent emission of the radiation, the resulting electric field can then be linearly scaled up to the measured charge of $131\,\mbox{pC}$.  This is a way to ``thin" the shower at a  $10^{-6}$ level.
The specifics of the implementation of the two formalisms are presented in the following subsections.


\subsubsection{Details of the implementation of the ZHS formalism}
The ZHS formalism calculates the radio emission of an individual particle track as a vector potential.  Details of the derivation can be found in~\cite{Alvarez-Muniz2010,PhysRevD.45.362,PhDAnne}.
 
Since a shower is considered to be a superposition of finite particle tracks (sub-tracks) with a constant velocity and the Coulomb field is negligible, the vector potential of a shower is simply given by the sum of the individual track-level vector potentials over all tracks for an observer position in the far-field.  The corresponding electric field is then given by the time-derivative.

\subsubsection{Details of the implementation of the endpoint formalism}

The implementation of the endpoint formalism in Geant4 was done in a way equivalent to the implementation in the CoREAS code~\cite{CoREAS}, computing the signal as an electric field in the time domain.  In the endpoint formalism, the electric field is calculated directly from the Li\'enard-Wiechert potentials.  However, rather than calculating the total emission from a track segment as is done in the ZHS formalism, the endpoint formalism considers the instantaneous acceleration of a charge at the beginning and end of the track segments to calculate the radio emission.  Details about the derivation of the endpoint formalism can be found in~\cite{James2011}, and further information on its implementation in the T-510 simulation can be found in~\cite{PhDAnne}.

The endpoint approach has the advantage that it does not rely on the Fraunhofer approximation (i.e., track segments need not be small with respect to wavelength and source distance), which might provide advantages in computational efficiency.
However, the calculation of the radio emission using the endpoint formalism becomes numerically unstable at the Cherenkov angle; here, a ZHS-like approach is used as a fallback. The threshold value for the fall-back depends on the medium and wavelengths of interest, i.e. it needs to be adapted for the calculation applied to an HDPE target.

To determine the appropriate threshold, the electric field from a single track with a length of $1\,\mbox{cm}$ at a distance of $10\,\mbox{m}$ to an observer is calculated as a function of the angle to the observer, $\theta$.  If $\theta$ comes close to the Cherenkov angle of about $49^\circ$, the electric field calculated using the endpoint formalism diverges and goes to infinity while the result for the ZHS formalism remains finite. This divergence is caused by the $1-n\beta\cos(\theta)$ term in the denominator of the endpoint formula (see Ref.~\cite{James2011}). For the simulation of the radio emission in the T-510 experiment, we set a threshold value of $(1-n\beta\cos(\theta))^{-1} = 10$.  This value safely excludes the singularity and yields stable results for the shower emission calculation which are not sensitive to variations of the threshold. The chosen threshold is equivalent to an observer angle of $\le5^\circ$ around the Cherenkov angle, within which the ZHS-like fall-back is used to calculate the expected radio signal for the corresponding tracks. (For air-shower simulations with CoREAS the threshold value is set to $1,000$.)

\subsection{Handling the velocity of the particle tracks}

A particle's sub-track velocity given by Geant4 represents the particle's velocity calculated at the beginning of its sub-track.  Due to the treatment of multiple scattering within Geant4, the end-point of the particle track gains a lateral displacement~\cite{GEANT}, which leads to an inconsistency between the position of the particle and its velocity and direction along the sub-track.  Therefore, the velocity and directional information provided by Geant4 are not directly usable for the calculation of the radio emission using the endpoint or the ZHS formalism.

Instead, these parameters have to be calculated on the basis of positions of sub-track start- and end-points and their corresponding times:
\begin{equation}
v_{\tiny{\mbox{sub-track}}}=\frac{|\vec{x}_{\tiny{\mbox{end}}} - \vec{x}_{\tiny{\mbox{start}}}|}{t_{\tiny{\mbox{end}}}- t_{\tiny{\mbox{start}}}}.
\end{equation}
However, this velocity calculated on the basis of positions and times reported by Geant4, can also exceed the velocity of light in vacuum of $c\approx 300\,\mbox{mm/ns}$.  This is because the sub-track gets longer due to the shift of the end-point, but the corresponding time $t_{\tiny{\mbox{end}}}$ is not adjusted consistently in the treatment of multiple scattering in Geant4.

To mitigate this problem, a maximum step length for each sub-track is chosen.  Reducing the length reduces the lateral displacement of the sub-track.  We found a value of $0.2\,\mbox{mm}$ to be an optimal sub-track length and resolved this problem for most particles. In the case of low energy particles, however, the effect of having a velocity along the sub-track higher than the speed of light is still observable.  Since the contribution from these particles to the total radio signal is expected to be negligible ($<1\%$), assuming they would follow a behaviour expected for a relativistic particle, these particles are skipped in the calculation by setting an energy cut of $E_{\mbox{kin}}> 0.1\,\mbox{MeV}$.

\subsection{Refraction and transmission effects}
\begin{figure}
\centering
\vbox{
\includegraphics[width=0.48\textwidth]{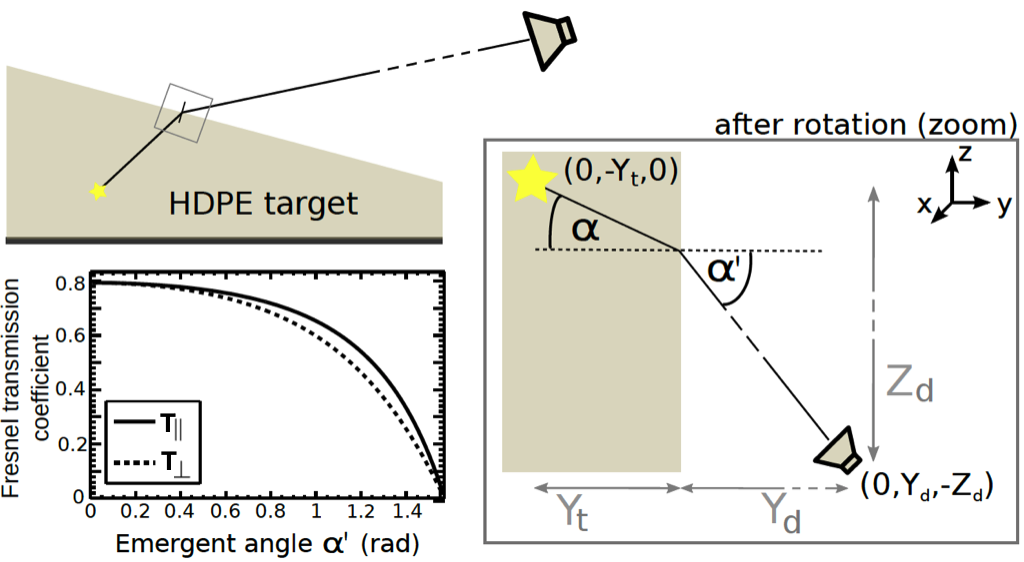} \\

\caption{Top, left: Sketch of the experiment geometry for the calculation of the point of refraction at the upper slanted target surface. Right: Enlarged view of the sketch after rotation to define the parameters for the calculation of the point of refraction.
Bottom, left: Values of the Fresnel transmission coefficient for the electric field components
parallel and perpendicular to the plane of incidence in dependency on the emergent angle. For further details see~\cite{Zilles2015}.}
\label{fig:Refraction}
}
\end{figure}

We consider refraction at the upper slanted target boundary as well as Fresnel transmission coefficients and demagnification effects~\cite{Lehtinien} in the propagation of the radio signals via ray optics.  The boundary conditions of Maxwell's equations dictate the change in the amplitude of an electric field passing trough a dielectric boundary. 
The ratio of the transmitted electric field to the incident one is given by $T = E_T/E_I $. Here, one has to distinguish two cases: the electric field parallel to the plane of incidence ($T_\parallel$) and the electric field perpendicular to the plane of incidence ($T_\perp$). The corresponding fraction of the electric field which is reflected 
at the boundary is defined by
\begin{align}
R_\mathrm{\parallel} &= \frac{n_\mathrm{Air}\cdot \cos{\alpha} - n_\mathrm{HDPE}\cdot \cos{\alpha'}}{n_\mathrm{Air}\cdot \cos{\alpha} + n_\mathrm{HDPE}\cdot \cos{\alpha'}}\\
R_\mathrm{\perp} &= \frac{n_\mathrm{HDPE}\cdot \cos{\alpha} - n_\mathrm{Air}\cdot \cos{\alpha'}}{n_\mathrm{HDPE}\cdot \cos{\alpha} + n_\mathrm{Air}\cdot \cos{\alpha'}}.
\end{align}
with $\alpha$ as the angle of incidence to the normal inside the target and $\alpha'$ as the angle of refraction (compare to Fig.~\ref{fig:Refraction}, top). 

Finally, the ratio  of the transmitted electric field to the incident electric field depends on the ratio of the refractive indices of both media and 
is given by the relation to the reflected part of the field as described in~\cite{Dawn}: 
\begin{align}
T_\mathrm{\parallel, \perp}=\sqrt{\frac{n_\mathrm{HDPE}}{n_\mathrm{Air}} (1-R_\mathrm{\parallel, \perp}^2)}.
\end{align}

The refraction of rays at a boundary between media with different refractive indices results in a shift in the apparent position of the radiation source. The treatment of the transmitted signal has to account for this.
The law of energy conservation requires that the incident power has to be equal to the sum of the reflected and transmitted power. 
On the assumption that the area $A$ on the target surface illuminated by the radiation is determined by a spherical wave front and that the angles of incidence and refraction, respectively, are constant in this area, 
the final analytical form for the Fresnel coefficients, including a correction for the spreading of the rays after refraction,  can be expressed analytically by~\cite{Dawn}:
\begin{align} 
T_\mathrm{\parallel} &= \sqrt{ \frac{\tan{\alpha}}{\tan{\alpha'}}(1-R_\mathrm{\parallel}^2)}
\label{eq:fresnelcoeff_par}\\
T_\mathrm{\perp} &= \sqrt{ \frac{\tan{\alpha}}{\tan{\alpha'}}(1-R_\mathrm{\perp}^2)}.
\label{eq:fresnelcoeff_perp}
\end{align}
Fig.~\ref{fig:Refraction}~(bottom, left) illustrates the Fresnel transmission coefficients' behavior as a function of the signal emergent angle. 
Larger emergent angles represent larger antenna heights with respect to the point of refraction, following a cosine-behavior.

The point of refraction on the upper surface has to be found individually for every combination of track and antenna position.
For the calculation of  the point of refraction, where the propagation time of the signal from the end-points of the sub-track to the antenna is the shortest and its path fulfills Snell's law, the coordinate system is transformed as shown in Fig.~\ref{fig:Refraction}~(right). Here, the target 
surface  is defined as the x-z-plane.
From geometrical considerations, the equation
\vspace{-0.1cm}
\begin{equation}\label{eq:refraction}
|Y_t|\cdot \tan{\alpha } = |Z_d| - |Y_d|\cdot{\tan \left(\arcsin \left(\frac{n_\mathrm{HDPE}}{n_\mathrm{Air}}\cdot \sin{\alpha}\right)\right)}
\end{equation}
can be derived, with $|Y_t|$ denoting the distance of the track to the surface and $|Y_d|$ the distance from the surface to the antenna. 
The parameter $|Z_d|$ represents the distance in $z$-direction to the point of refraction and
$\alpha$ the angle of incidence.
In the simulation all parameters are known. They can be used to calculate the point of refraction analytically. Equation~\ref{eq:refraction} is used for every track and antenna combination, 
requiring that the line of sight intersects at the upper target surface. 
Once the point is known, it is possible to calculate the incident and the emergent angle to the normal of the upper target surface and the corresponding Fresnel transmission 
coefficients for the vertical and horizontal polarization components of the 
electric field with respect to the plane of incidence. This leads to the transmitted electric field
\vspace{-0.1cm}
\begin{equation}
\vec E_{\mathrm{ant}} = (\vec E_{\mathrm{em}} \cdot \hat{\vec r}_{\perp,\, \mathrm{in}} ) \cdot T_\perp \cdot \hat{\vec r}_{\perp,\, \mathrm{out}} + (\vec E_{\mathrm{em}} \cdot \hat{\vec r}_{||,\, \mathrm{in}} ) \cdot T_{||} \cdot \hat{\vec r}_{||,\, \mathrm{out}}.
\end{equation}
The parameter $\vec E_{\mathrm{em}}$ is the electric field emitted by the track, $\hat{{\vec r}}_{\perp}$ the vector which is perpendicular to the plane of incidence ($in$) 
as well as after being refracted at the boundary ($out$) and 
$\hat{{\vec r}}_{||}$ is the vector lying in the plane. 
The factors $T_{\perp}$ and $T_{||}$ are the corresponding Fresnel transmission coefficients~\cite{Lehtinien} 
for the perpendicular and parallel polarization components, respectively.  The time $t_{prop}$ for the signal propagation from track to antenna is obtained directly from using the point of refraction and is given by:
\vspace{-0.1cm}
\begin{equation}
t_{prop} = n_\mathrm{HDPE} \cdot \frac{d_\mathrm{HDPE}}{c} + n_\mathrm{Air} \cdot \frac{d_\mathrm{Air}}{c}
\end{equation}
with the distance between track and point of refraction $d_\mathrm{HDPE}$ and the distance between point of refraction and antenna $d_\mathrm{Air}$.
The time $t_{prop}$ has then to be added to the time when the signal is emitted at the track.

\subsection{Cherenkov-like effects reproduced by the simulation}

\begin{figure}
\centering
\vbox{
\includegraphics[width=0.45\textwidth]{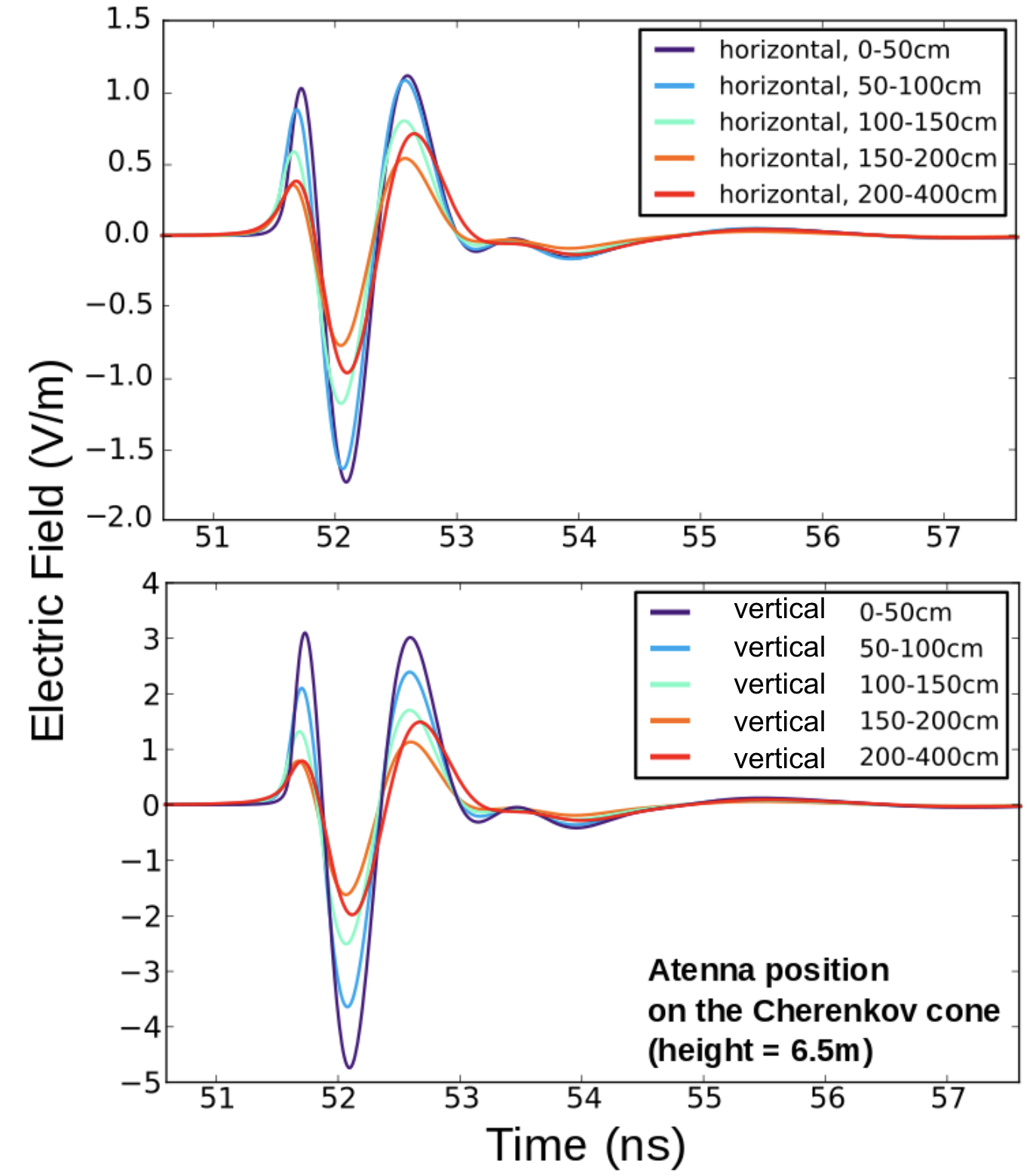} \\

\caption{Testing Cherenkov-like effects using the endpoint formalism: Horizontally (top) and vertically polarized (bottom) components of the electric field produced by a ``sliced'' particle shower in a magnetic field with a strength of $B=-970\,\mbox{G}$ for an antenna on the Cherenkov cone, using the endpoint formalism and filtered from $300-1200\,\mbox{MHz}$.}
\label{fig:CHERENKOV}
}
\end{figure}

Emission is enhanced at the Cherenkov angle because radiation emitted from the entirety of the particle shower arrives simultaneously, compressing the emission in time.  In order to demonstrate that this effect is seen in simulations, the radiation emitted along different tracks of the shower was calculated.  The tracks correspond to distances 0--50~cm, 50--100~cm, 100--150~cm, 150--200~cm, and 200--400~cm inside the target.  Fig. \ref{fig:CHERENKOV} shows the contributions from each slice for horizontally and vertically polarized emission based on the endpoint formalism for an antenna position on the Cherenkov cone.  It is clear that at this position the contributions from different slices add up coherently.  Additionally, we see that most of the contributions for the vertically polarized signal come from the first 50~cm of the target, while the horizontally polarized signal has similar contributions from both the 0--50~cm and 50--100~cm tracks.  This is consistent with the design of the experiment.  The shower begins in the lead target, and so the charge-excess component of the radiation, which is measured in the vertical polarization, begins early.  The magnetically-induced component, only seen in the horizontal polarization, begins when the shower enters the magnetic field, and so develops later in the target.

\section{Application of the simulations to the T-510 experiment}

The electric field in the time domain has been calculated using the endpoint formalism for a 2D grid of antenna locations in the x-z plane (see Fig.~\ref{fig:T510}) with positions in $0.5\,\mbox{m}$ steps and with a primary electron energy of $4.35\,\mathrm{GeV}$. The horizontal distance to the entry point of the electrons in the target is about $13\,\mbox{m}$.  We use magnetic field values from the measured 3D magnetic field map in the Geant4 simulations, which have the maximum strength of up to $970\,\mbox{G}$ in the vertical direction for an applied current of $2400\,\mbox{A}$ in the vertical direction perpendicular to the electron beam.
 
\begin{figure}
\includegraphics[width=0.48\textwidth]{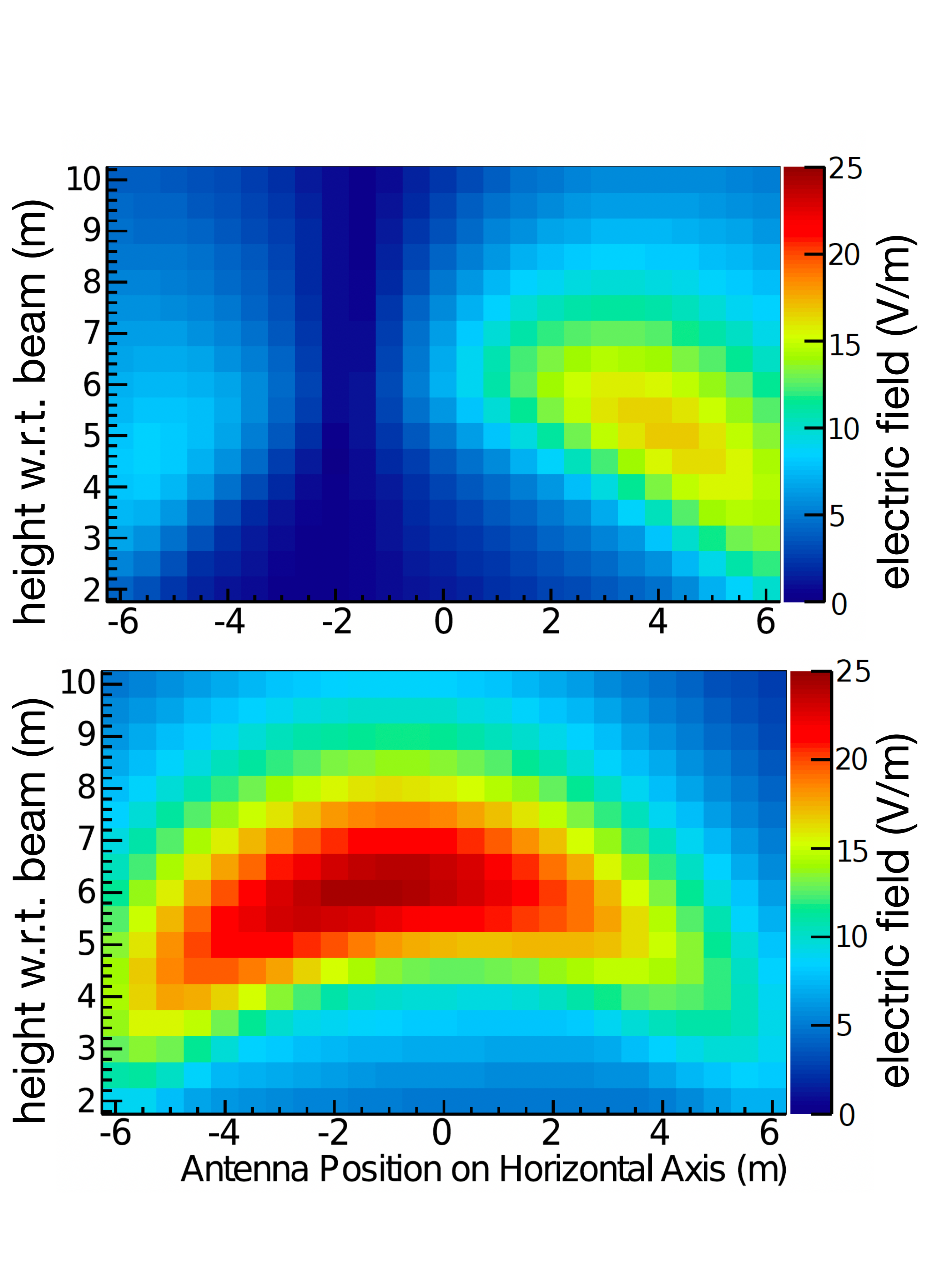}
\caption{Peak amplitude of the electric field for a 2D antenna array using the endpoint formalism for a magnetic field of a maximum 
strength of $970\,\mbox{G}$: Top: Horizontally polarized component. Bottom: Vertically polarized component~\cite{Zilles2015}.}
\label{fig:2D_total_B_end}  
\end{figure}

The peak amplitudes of the horizontally and vertically polarized electric fields for this 2D map are shown in  Fig.~\ref{fig:2D_total_B_end}.   The positions of the maximum values for the peak amplitude of the signal form a strong Cherenkov ring, whose position agrees with the expectation given by the refractive index 
of $n_{\text{HDPE}}=1.53$ and $n_{\text{Air}}=1.0003$.  The finite target size leads to a cut-off of the Cherenkov ring on both sides. The elliptical appearance of the ring is caused by the refraction at the slanted target surface.   Asymmetries can be seen off axis, and are primarily due to interference between magnetically-induced and Askaryan radiation. Ideally, on the x=0 axis, the magnetically-induced and Askaryan radiation would be confined to independent polarizations. From the measured magnetic field maps (Fig.~\ref{fig:Bmap}) non-vertical components can be seen in the magnetic field. These components contribute to a tilting of the shower development, introducing additional asymmetrical effects. A realistic magnetic field is used in the simulations, and so these effects are accounted for in the simulated results which will be compared to data.  

\subsection{Comparison of simulation results using the endpoint and the ZHS formalisms}
\begin{figure}
\centering
\includegraphics[width=0.46\textwidth]{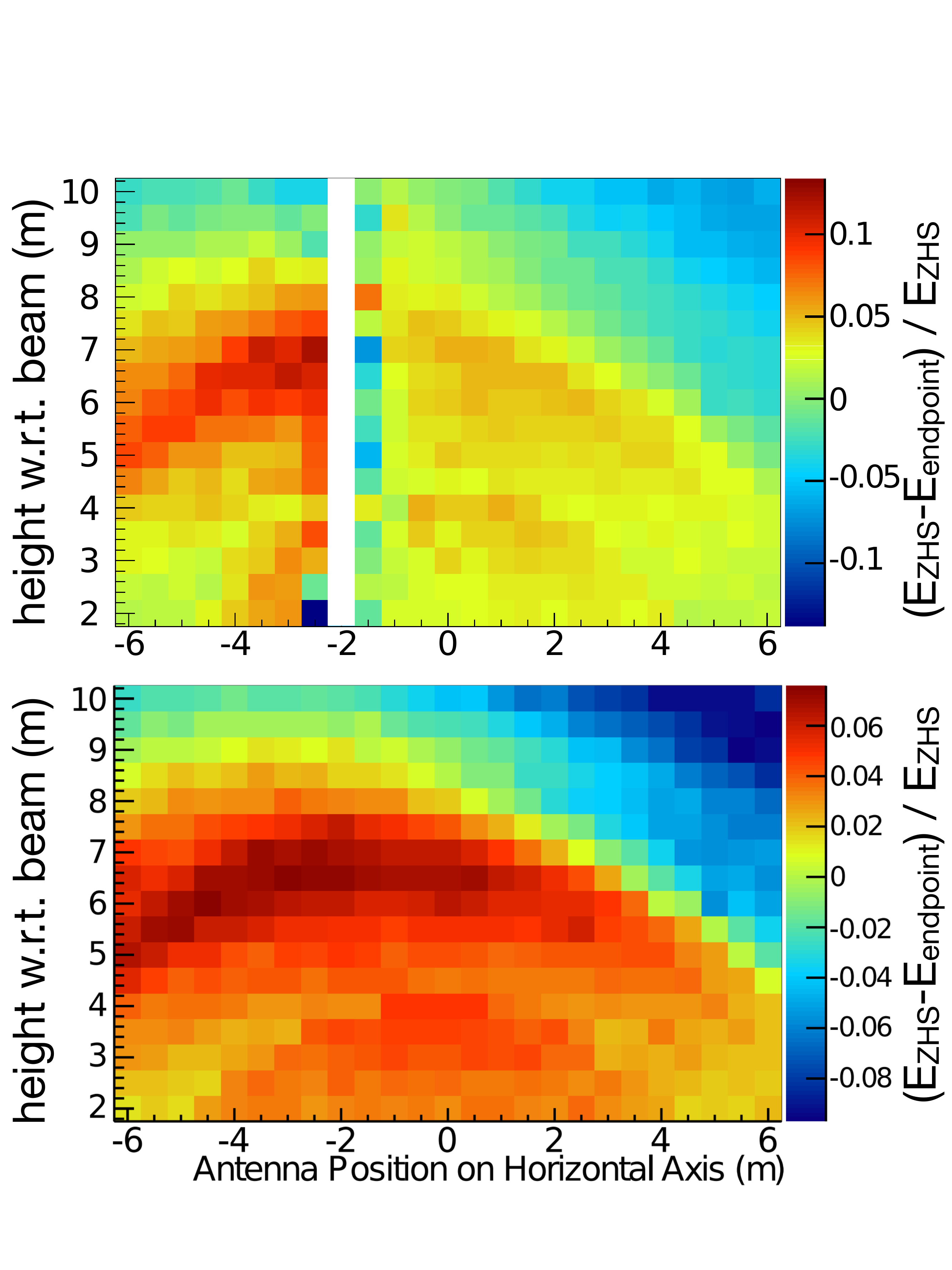}
\caption{Relative deviation of the peak amplitude in the time domain defined as $\frac{E_\mathrm{ZHS}-E_\mathrm{endpoint}}{E_\mathrm{ZHS}}$ for a magnetic field strength of $B=970\,\mbox{G}$: Top:  Horizontally polarized component. Antenna positions with negligible signals have been excluded in the comparison (white area). Bottom: Vertically polarized component ~\cite{Zilles2015}.}
\label{fig:Diff}
\end{figure}
Since the calculation of the radio signal from a particle shower can be done in parallel with both formalisms, we can perform a direct comparison of their results to study possible differences due to numerical aspects and approximations underlying the formalisms. 
Fig.~\ref{fig:Diff} shows the 2D distribution of the relative deviation of the peak amplitude in the time domain between the endpoint formalism and the ZHS formalism for the maximum magnetic field strength of $970\,\mbox{G}$ using the realistic field map and a primary beam energy of $4.35\,\mathrm{GeV}$. 
A ring structure is still visible in the distribution. Despite deviations up to the~10\% level, no systematic offset between the results of the two formalisms can be observed. However, the comparison shows that the deviations in the horizontally (top) and vertically polarized components (bottom) depend on the position of the antenna.  Since there is little horizontal radiation at the position $x=-2$~m, the ratio diverges at this location. 

This leads to the conclusion that the formalisms reproduce the contributions due to the magnetic effect in a slightly different ways. Furthermore, inside the Cherenkov ring the ZHS formalism leads to slightly higher results and the endpoint formalism predicts slightly higher amplitudes outside the ring. The origin of these differences has to be studied in more detail, which will be facilitated by the inclusion of both formalisms in CORSIKA~8~\cite{Karastathis:20218X}, for example.


\subsection{Transition radiation}\label{TR}
\begin{figure}
\includegraphics[width=0.46\textwidth]{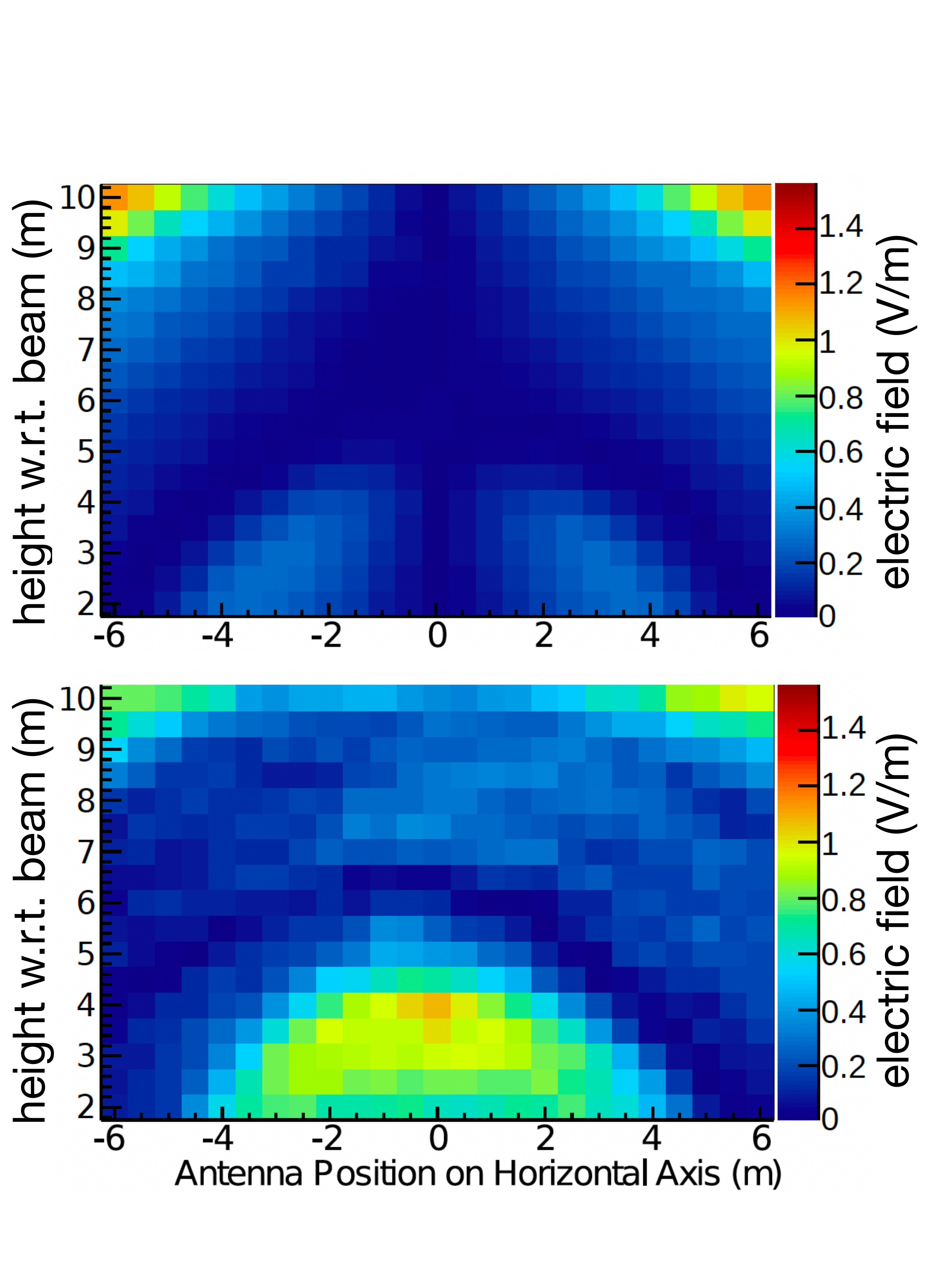}
\caption{Contribution of transition radiation to the peak amplitude in the time domain. Top:  Horizontally polarized component.  Bottom: Vertically polarized component.  The  relative  contribution  by  the  transition  radiation to the radio signal strength is about 1\% for both polarizations near the Cherenkov cone.}
\label{fig:2D_TR}
\end{figure} 
Charged particles crossing a boundary of media with different refractive indices produce transition radiation, in this case at the boundary of the lead pre-shower medium and the HDPE target. This leads to a possible additional source of radiation which can be estimated with Geant4 simulations using the endpoint formalism.  The steps in the simulation program are limited by the boundary of the current volume. This means that the step ends exactly at the boundary of the lead.  The electric field produced by steps in the lead can not escape. The following steps of the particle track start directly in the HDPE target at the boundary.  The electric field produced at its start point represents the contribution to the signal from the expected transition radiation~\cite{James2011}.

The absolute contribution of the transition radiation to the signal of the whole shower for a 2D antenna array at about a horizontal distance of $13\,\mbox{m}$ to the entry point of the beam is shown in Fig.~\ref{fig:2D_TR}. Since the magnetic field already starts to affect the shower during the pre-shower stage, a small asymmetry in the signal distribution is observable.

The relative contribution by the transition radiation to the radio signal strength is of order $1\%$ for the horizontally as well as for the vertically polarized components for antenna position close to the Cherenkov angle~(compare the absolute scale to Fig.~\ref{fig:2D_total_B_end}), the impact of transition radiation is thus negligible for this study.

\section{Comparison of the simulation results to measured data}

\subsection{Convolving the simulations with the detector response}

In order to directly compare the simulation results from the endpoint and ZHS formalisms with the measured voltages, we convolved the simulations with the measured system response of the cables and filters for each channel and the impulse response of the antennas.

At the antennas, the voltage in the time-domain is described by the convolution
\begin{equation}
 V(t) = h_{\text{eff}}(t) \circ h_{\text{sys}}(t) \circ E(t) 
\end{equation}
with the antenna impulse response $h_{\text{eff}}(t)$, the system impulse response due to filters and cable losses $h_{\text{sys}}(t)$ and the simulated electric field $E(t)$. This is equivalent to the multiplication of the effective height, $h_{\text{eff}}(f)$, the system response 
$h_{\text{sys}}(f)$ and the electric field, $E(f)$, in the frequency domain.  In addition the simulations are down-sampled to $10\,\mbox{GSPS}$ while the measured data are upsampled to match.  Further details of the data processing can be found in~\cite{StephICRCProc}.

\subsection{Systematic uncertainties}

In this section we address the known systematic uncertainties in the experiment as they effect the pulse amplitude.  These include beam charge calibration, magnetic field strength, and antenna geometry. The beam charge measurements yield a 2\% systematic uncertainty overall in a charge bunch. 
The magnetic field was monitored at the same point in the target for all runs, resulting in a root-mean-squared variation of 72 Gauss for full strength field. This adds a 6\% uncertainty.
The antenna array was adjusted manually with ropes for each measured position and the height was determined with a laser measure.  The largest geometric uncertainty is in the antenna angle relative to the target, and is estimated to be 6\%, as the antenna response does not change significantly for angular differences below $10^\circ$.  

The known systematic uncertainties in simulations are due to the difference in peak amplitude between endpoint and ZHS formalisms, the assumption of ray optics, and the validity of assuming the antennas are in the far field of the emission from the target when calculating the transmission coefficients.  These contributions to the systematic uncertainties are summarized in Table~\ref{table:systematic_errors}.  There is also an uncertainty due to the reflection coefficient of the RF absorbing blanket beneath the target listed in Table~\ref{table:systematic_errors}.  This uncertainty has a 6\% effect on the pulse amplitude and will be discussed in detail in the following sections.

\begin{center}
\begin{table}
\centering
 \begin{tabular}{||c c c||} 
 \hline
 & \textbf{uncertainty} &   \\ [0.5ex] 
 \hline\hline
  \textbf{Simulation} & ZHS vs. endpoint  & 5\% \\
   \hline
 &  propagation & 5\%\\
 \hline
 &  reflection coefficient & 6\%\\
 \hline
 \textbf{Total} &  & \textbf{9.3\%}\\

  \hline\hline
 \textbf{Data}  &  beam charge & 2\%\\
 
  \hline
 & magnetic field& 6\%\\
 
  \hline
 & antenna alignment & 6\%\\
 
  \hline
  \textbf{Total} &  & \textbf{8.7\%}\\

\hline
\end{tabular}
\caption{Summary of systematic uncertainties on the T-510 data and simulations.}
\label{table:systematic_errors}
\end{table}
\end{center}

\subsection{Comparison of the simulated and measured radio signal at the Cherenkov angle}

The agreement of the absolute scale of the simulated and measured radio emission was previously studied in~\cite{Belov2016}.  Here we briefly summarize the results and limitations of the original study.

The peak amplitude of each time-domain signal is taken as the value to compare between experimental data and simulation.  This quantity is chosen because it is straight forward to determine, and is more stable than the power in the trace, which is more influenced by ringing due to filters and reflections in the target.  We chose the antenna position closest to the Cherenkov ring at a height of $6.5\,\mbox{m}$. At this position, the vertical channel of the antenna measures the Askaryan component of the radio signal and the horizontal channel measures the magnetically-induced component of the radio signal, so each effect can be studied separately. 

As already illustrated in Fig.~\ref{fig:Diff}, the peak amplitude predicted by the two formalisms agrees within $4.1\%$ in the horizontally polarized component and within $4.9\%$ in the vertically polarized component for this antenna height. This leads to the conclusion that the two formalisms deliver consistent results at that position.  From the work shown in~\cite{Belov2016} we saw that although the simulations could reproduce the shape of the measured signal well, the peak amplitude of the data  exceeded the simulations in both polarizations by about $35\%$ in the time domain.  We discuss the origin and resolution of this discrepancy in the following section.

\subsection{Including internal reflections}\label{handelreflctions}

\begin{figure}
\centering
\includegraphics[width=0.45\textwidth]{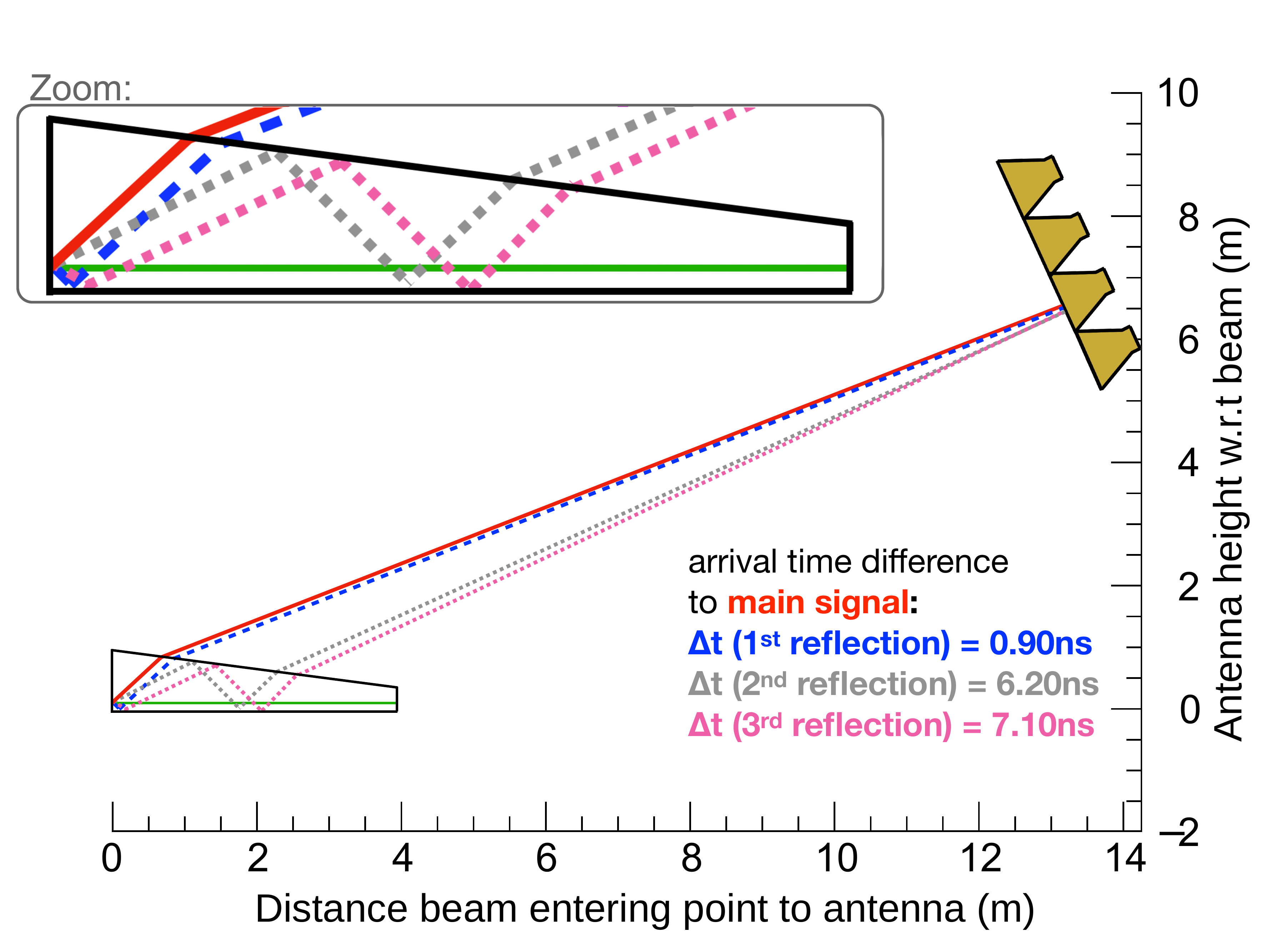}
\caption{Ray tracing of the main signal and internal reflections that reach the same antenna.}
\label{fig:reflectionTarget}
\end{figure}

The discrepancy between the simulated results and the measured data is primarily due to internal reflections from the interface of the bottom surface of the target and the RF absorbing blanket beneath it.  Although it is possible that reflections off the back of the target contribute to the signal, we only consider three reflections from the bottom of the target which are shown schematically in Fig.~\ref{fig:reflectionTarget}.  The first reflection (in blue) is separated from the main signal by about 1~ns in time. The second reflection (in grey) and third reflection (in pink) are separated from the main pulse by close to 6~ns.  Although lower in amplitude, they create an interference pattern visible in the frequency domain, with a beating every 150~MHz.

\begin{figure*}
\centering
\includegraphics[width=\textwidth]{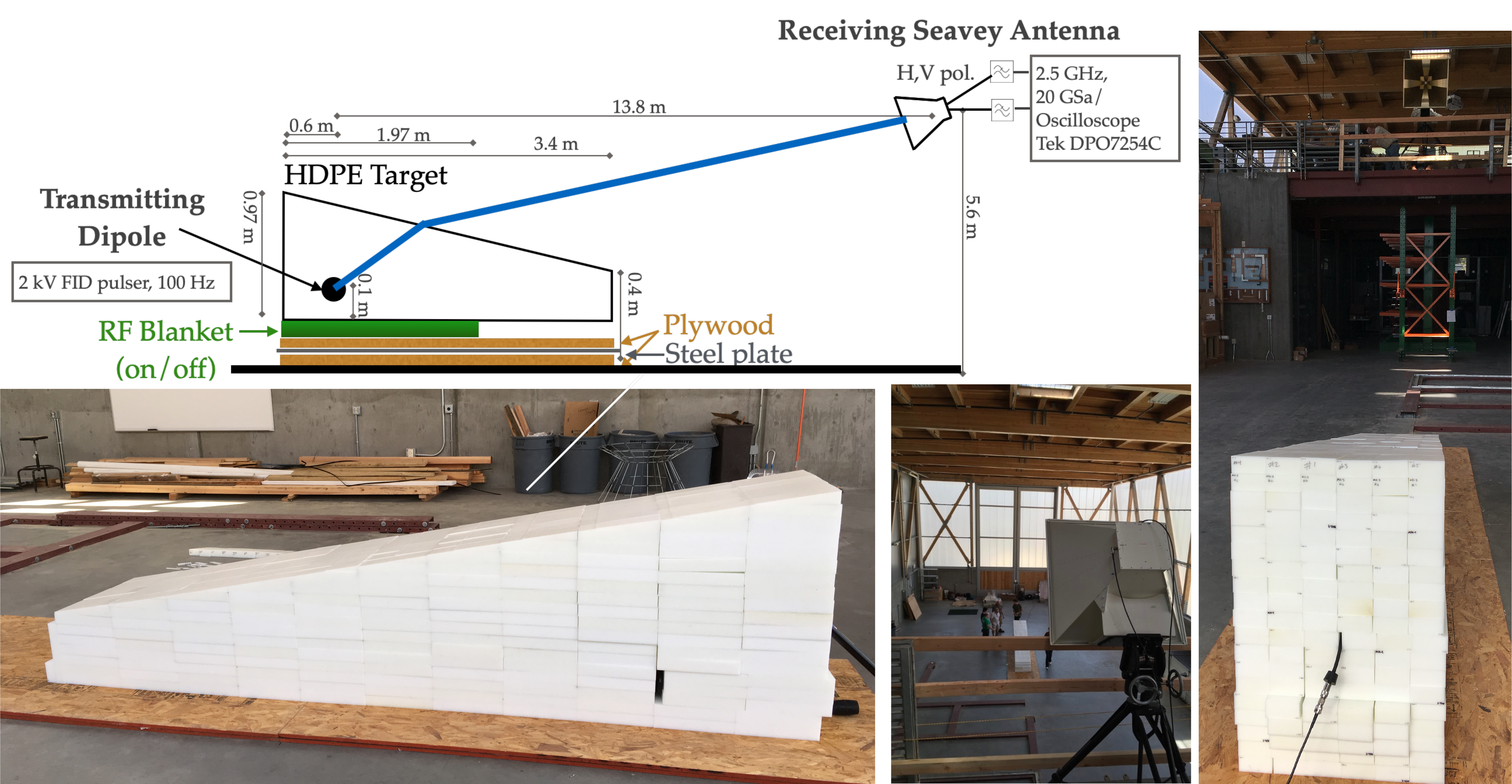}
\caption{The CP-510 experiment designed to measure the reflection coefficient of the RF blanket and plywood used in the original T-510 experiment. The primary difference between the two runs used to measure the reflection coefficient is that the RF blanket was removed between the runs.}
\label{fig:cp-510}
\end{figure*}
At the time of the experiment, the frequency-dependent reflection characteristics of the target-to-blanket boundary below 6~GHz were not specified, and so an implementation of the reflection in the simulation was not possible.  A follow-up experiment was conducted to measure the unknown reflection coefficient.  

The experimental design for the reflection measurement, named CP-510, is shown in Fig.~\ref{fig:cp-510}. The HPDE target from the original T-510 was re-assembled in the Simpson Strong Building at the California Polytechnic State University on top of two sheets of 1/2-inch thick plywood with a 1/4-inch steel plate between them. A Telewave 400D folded dipole was installed 0.6~m from the end of the target.  The dipole was aligned parallel to the horizontal. The nulls of the embedded dipole pointed towards the narrow sides of the target. The dipole was dielectrically loaded by the HDPE such that its measured reflected power was less than 25\% (\textit{i.e.} $S_{11} < -6$ dB) between 265 and 1300~MHz.  The transmitting dipole was driven by a 2~kV FID pulser at 100~Hz repetition rate. The signal was received by a quad-ridged horn antenna (similar to the ones used for T-510) situated on a mezzanine above the target.  

\begin{figure}
\centering
\includegraphics[width=0.45\textwidth]{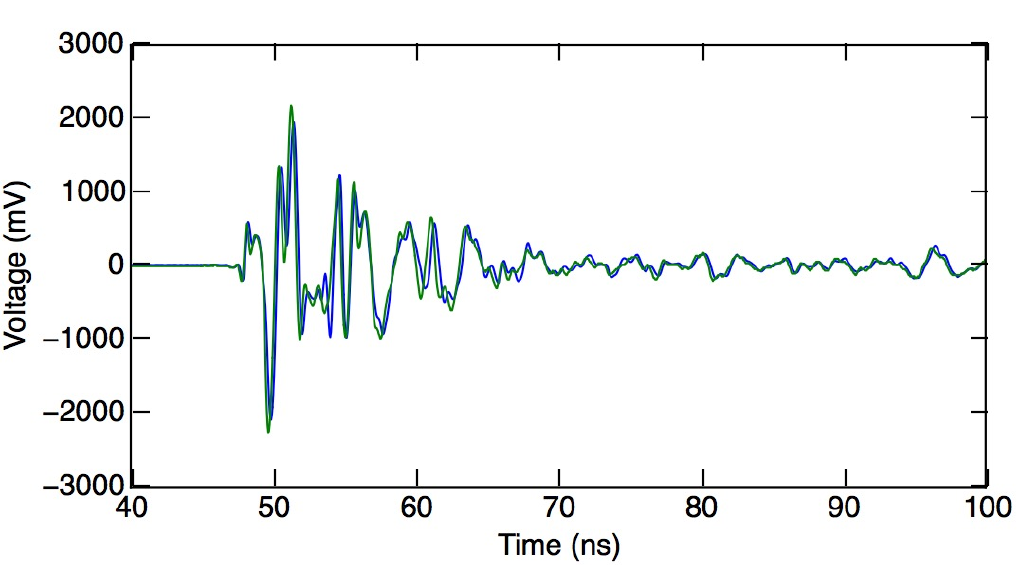}
\caption{Waveforms measured during the follow-up reflection experiment, CP-510, from the co-polar feed (Hpol) on the horn antenna. The pulses are shown for cases both with (green) and without the RF blanket (blue) on top of a steel reflector. The pulses are nearly identical indicating that the blanket acts primarily as a reflector rather than an absorber at these frequencies.}
\label{fig:cp-510-pulses}
\end{figure}

The signal transmitted from the embedded dipole to the receiver was recorded in two runs, with and without the RF absorbing blanket. For the runs with the blanket, the blanket was placed between the target at the topmost piece of plywood. This was done to closely mimic the setup at T-510, where the RF absorber was placed atop plywood and the magnetic field coils, which were both on top of a steel plate. Thus the CP-510 experiment measures the reflection coefficient of the blanket and the plywood, but neglects the magnetic field coils that were present in T-510. We took care to align the receiving antenna with the direction of the signal from the transmitting antenna. For each run, an oscilloscope recorded an average of 500 events. The measured impulses in the co-polar (horizontal) direction for runs with and without the blanket are shown in Fig.~\ref{fig:cp-510-pulses}.

The waveforms measured with and without the RF blanket are nearly identical, indicating that the blanket does not absorb radiation at the relevant frequencies. The absorbing material of the mats is rated only for frequencies above 1~GHz, and indeed destructive testing confirmed that the material contained a layer of conductive mesh that would destructively degrade a signal at higher frequencies. However, we can use these measurements to estimate the reflection coefficient of this material.

\begin{figure}
\centering
\includegraphics[width=0.45\textwidth]{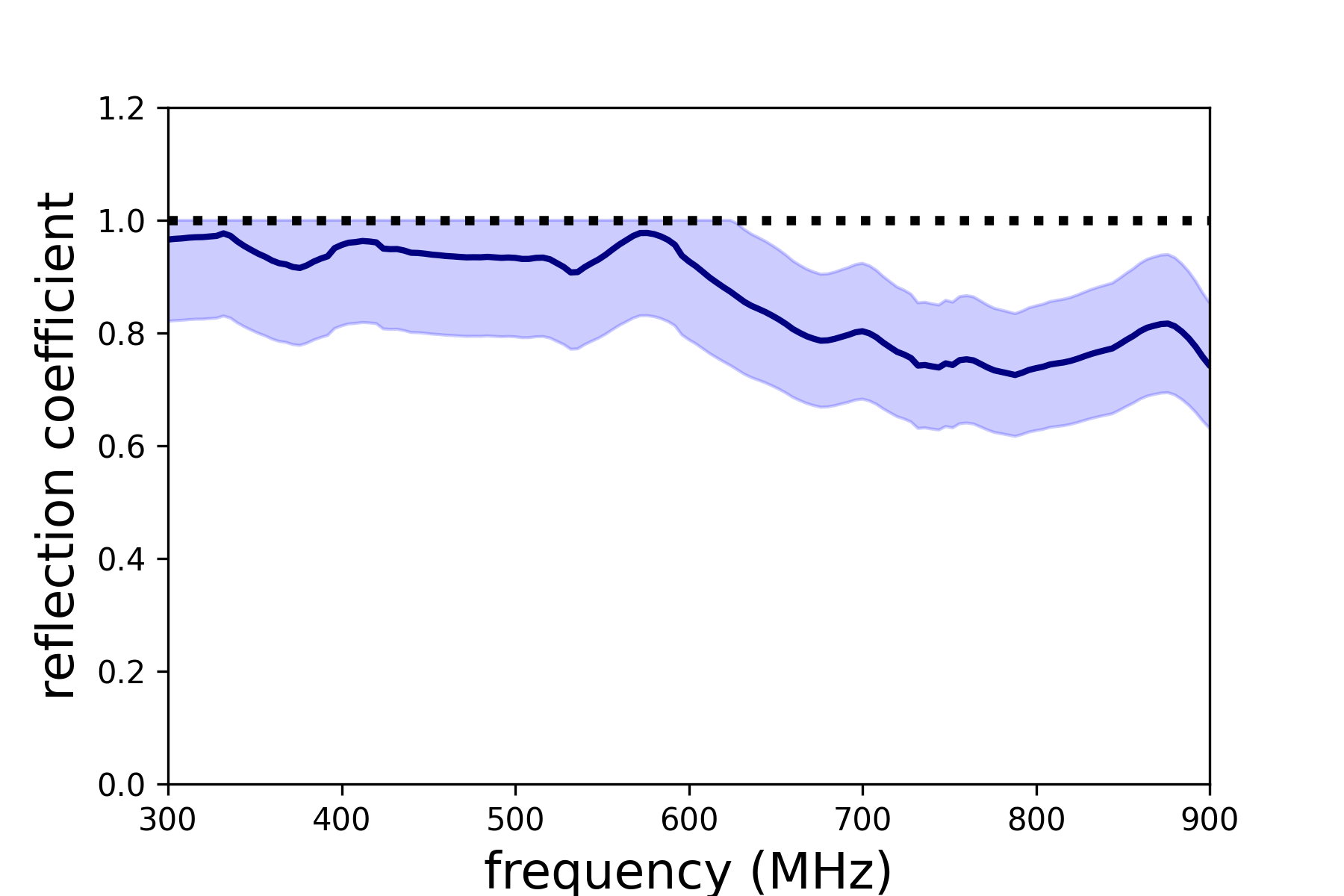}
\caption{Measured reflection coefficient of the RF  blanket.  We use this as the reflection coefficient for both polarizations.  The shaded area reflects the uncertainty, which is $\pm$15\% with an upper limit on the allowed reflection coefficient of 1.}
\label{fig:reflection_coeff}
\end{figure}

The reflection coefficient shown in Fig.~\ref{fig:reflection_coeff} is the ratio of the measured voltages with and without the blanket in the frequency domain.  We use the $300-900$~MHz band as that is where the measurement was most stable.  The setup included additional interference effects due to multiple reflections in the room and within the target that are different than the original setup in T-510. This can result in constructive interference which could artificially increase the reflection coefficient, sometimes even above $1$, especially because the mat acts as a good reflector. It can also result in destructive inference causing dips in the reflection coefficient. Because we expect the reflection coefficient to be smoothly varying, we smoothed the results in the frequency domain with a 50 sample rolling average, and we bound the uncertainty on the reflection coefficient by the percent difference between unsmoothed and smoothed cases. This results in a 15\% uncertainty in the reflection coefficient across the band,  and we restrict the upper limit to a maximum reflection coefficient to 1.  This uncertainty propagates into a maximum 6\% difference in observed signal amplitude, as indicated in Table~\ref{table:systematic_errors}.

\begin{figure}
\centering
\includegraphics[width=0.4\textwidth]{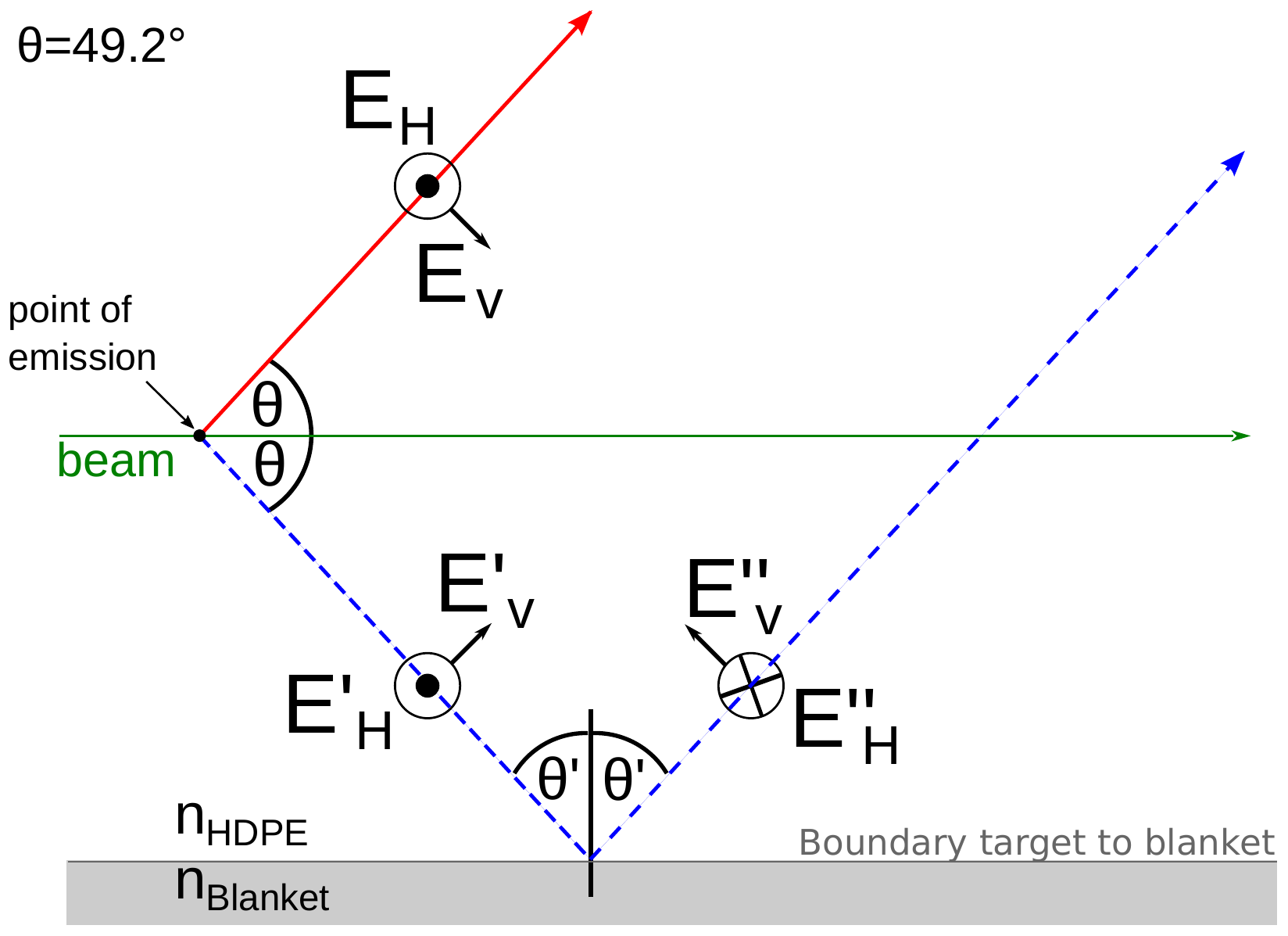}
\caption{Illustration of the separation of the electric field into a horizontally polarized component (E$_{\text{H}}$) and a vertically polarized component (E$_{\text{V}}$). It is also shown that the horizontally polarized component experiences a phase shift of $\pi$ at the boundary of the target to the RF absorbing blanket.}
\label{fig:Sketchreflection}
\end{figure}

With a known reflection coefficient, we take the approach of adding the reflections to the simulations and making comparisons in the $300-900$~MHz band.  As the convolution of the simulated signal with the detector response is a linear operation, we first convolve the direct signal and reflected signals separately and then sum them.  This requires knowing the shape, time delay, and amplitude of each reflected pulse.  The time delay of each reflection is calculated directly using known geometry.  Since most of the radiation arriving at the antenna comes from the beginning of the target, the entry point of the beam in the HDPE target is chosen as the reference point for the ray tracing.  To find the shape of the reflected signals we start at the antenna position and use ray tracing to find the angle of emission for the ray that would start at the same place as the direct emission, reflect, and hit the antenna. Then, with that emission angle we find the location at which a direct ray would land in the antenna plane.  Simulated data is only available for discrete antenna positions, so we use the simulated trace for the antenna position closest to the predicted location. For the first reflection this is the same antenna as for the main signal. For the second and third reflections this emission angle is offset by approximately $20^{\circ}$, twice the tilt of the target surface.  If the second and third reflections were to escape at the point where they first hit the top of the target, they would be seen by the bottom antenna, and so we use the signal shape simulated for the lowest antenna position for this reflection.
\newline \indent The amplitude of each reflection is determined using the measured reflection coefficient of the blanket and transmission coefficient at the top of the target.  The transmission coefficient at the upper surface of the target is recalculated to take into account the new geometry for each reflection.  We also consider a polarity flip at the target-blanket surface.  This geometry is shown in Fig.~\ref{fig:Sketchreflection}. 

The reflection coefficient and the flip of the sign for the horizontal component ($E_{\text{{H}}}'$ to $E_{\text{{H}}}''$) is given by the corresponding Fresnel coefficients. For a non-magnetic, dielectric medium the Fresnel coefficient reduces to~\cite{Jackson}:
\begin{equation}
 \frac{E_{\text{{H}}}''}{E_{\text{{H}}}'}= \frac{n_1 \cos(\theta') -\frac{\mu_1}{\mu_2} \sqrt{{n_2}^2 - {n_1}^2 \sin(\theta')} }{ n_1 \cos(\theta') +\frac{\mu_1}{\mu_2} \sqrt{{n_2}^2 - {n_1}^2 \sin(\theta')} }
 \end{equation}
 \begin{equation}
 = \frac{n_1 \cos(\theta') - \sqrt{{n_2}^2 - {n_1}^2 \sin(\theta')}}{n_1 \cos(\theta') + \sqrt{{n_2}^2 - {n_1}^2 \sin(\theta')}}.
\end{equation}
With $\mu_1 = \mu_2= \mu_0$, assuming a refractive index of the blanket which is larger than the one of the target ($n_1=n_{\text{\tiny{HDPE}}}$ and $n_2 = n_{\text{\tiny{blanket}}})$ and an incoming angle of $\theta'=40.8^\circ$ to the normal of incidence returns a ratio of the incoming signal to the reflected one of $-1 \geq\frac{E_{\text{\tiny{H}}}''}{E_{\text{\tiny{H}}}'} \ge 0$. 
A similar calculation can be done for the vertical component of the emitted signal. It returns a ratio of $0 \leq \frac{E_{\text{\tiny{V}}}''}{E_{\text{\tiny{V}}}'}\leq 1$. 
Note that the vertical component acquires a phase shift of $\pi$ for the emission towards the bottom surface just by geometry, as shown in Fig.~\ref{fig:Sketchreflection}.

\begin{figure*}
\centering

\includegraphics[width=1.0\textwidth]{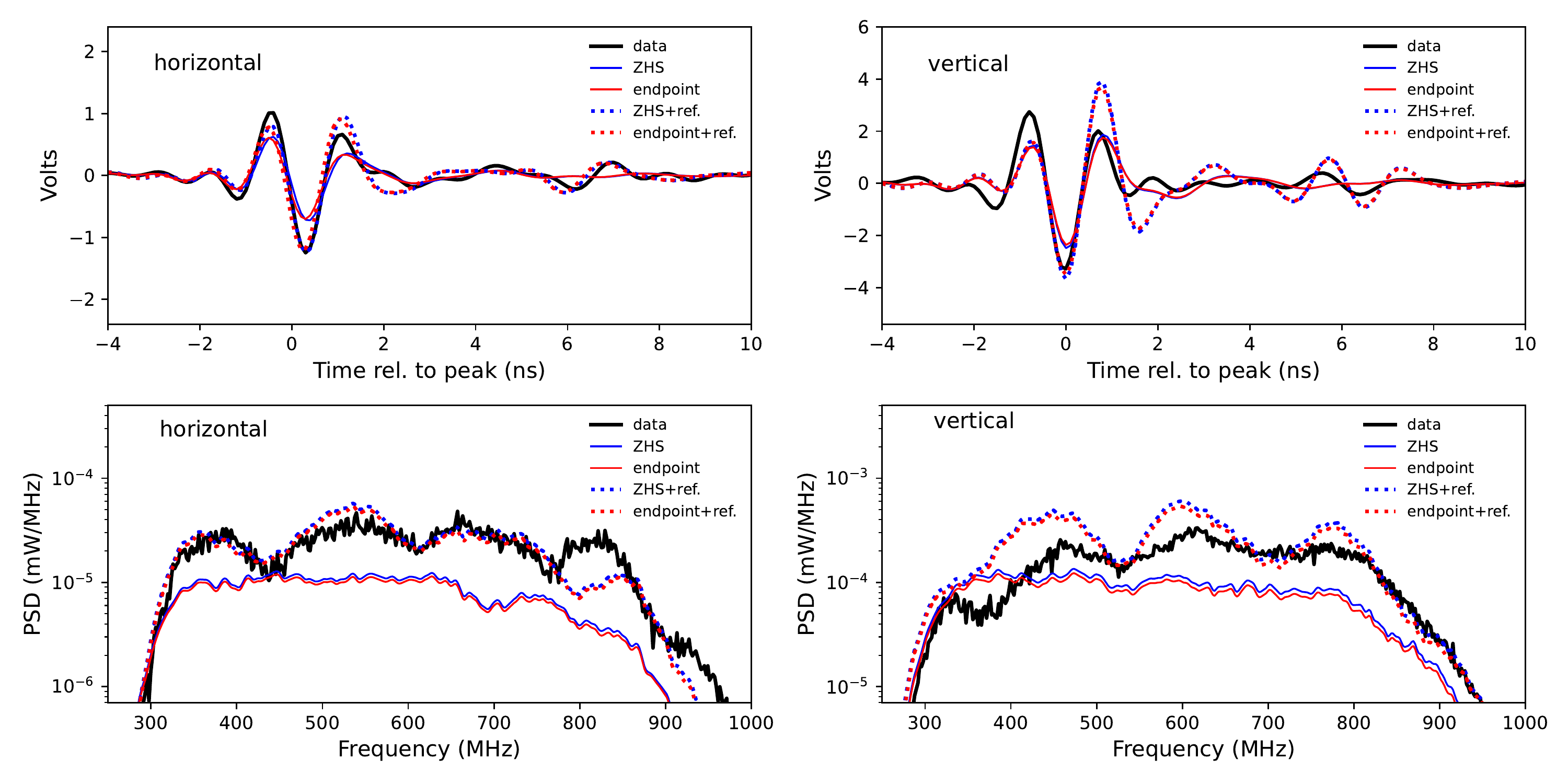}

\caption{Horizontal (left) and vertical (right) polarization for an antenna position at $6.5\,\mbox{m}$ height: Simulated time traces (top) and frequency spectra (bottom) are shown for data (in black) and for endpoint and ZHS simulations including reflections, in blue and red respectively. Simulations without reflections are shown as solid lines and simulations with reflections are shown as dashed lines.}
\label{fig:withreflection}
\end{figure*}

Fig.~\ref{fig:withreflection} shows the results of including reflections in the simulations.  Data measured by an antenna close to the Cherenkov cone is compared with the corresponding simulations 
(including reflections), based on the endpoint and the ZHS formalisms, shown as red and blue dashed lines.  The simulations without reflections are shown as solid lines.  We compare the amplitude of the down-going pulse.  Including the reflections we can describe the measured peak amplitudes very well for both polarizations.  For the vertically polarized component, the measured pulse peak is about 5\% lower than the corresponding simulation including the reflections in the case of the endpoints formalism and $11\%$ for the ZHS formalism.  For the horizontally polarized component the agreement is even better. The data exceeds the resulting peak amplitudes of the simulations based on the endpoint formalism by less than $5\%$, and based on the ZHS formalism by less than $1\%$.   The difference in the simulated peak amplitude between the two formalisms is $5\%$ which does not differ from the simulated results without inclusion of reflections.

Furthermore, the interference pattern seen in the data as a result of the reflections can also be reproduced.  The exact timing of the reflections determines where the points of maximum interference occur, and we use the timing as determined in the diagram in Fig.~\ref{fig:reflectionTarget} for both polarizations.  However, the different types of radiation (magnetically-induced and Askaryan) develop differently in the target.  The magnetically-induced radiation occurs primarily once the shower has exited the lead and experiences the magnetic field induced by the coils.  The fact that the timing of the peak emission is slightly different for the two polarizations could be the reason that the assumed reflection timing works better in one case than the other.  Additionally, the vertically polarized emission, which develops earlier, may be more affected by reflections off of the back of the target which are not included in the simulations.  Nevertheless, we conclude that including internal reflections from the bottom of the target sufficiently resolves the discrepancy between measurements and simulation.

\subsection{Magnetic-field scaling of the radio emission}

The strength of the magnetically-induced radio emission from a shower is expected to scale linearly with the magnetic field.  In contrast, the strength of the Askaryan emission should not be affected by the magnetic field. The T-510 experiment was able to demonstrate this effect in the lab by changing the magnetically-induced, horizontally polarized, component of the radio signal via manipulation of the applied magnetic field strength.  Since the magnetically-induced and charge excess components of the signal were separated into different polarizations, the ratio of the peak amplitudes measured in the horizontal and vertical channels, $H/V$, is studied. 

\begin{figure}
\centering
\vbox{
\includegraphics[width=0.48\textwidth]{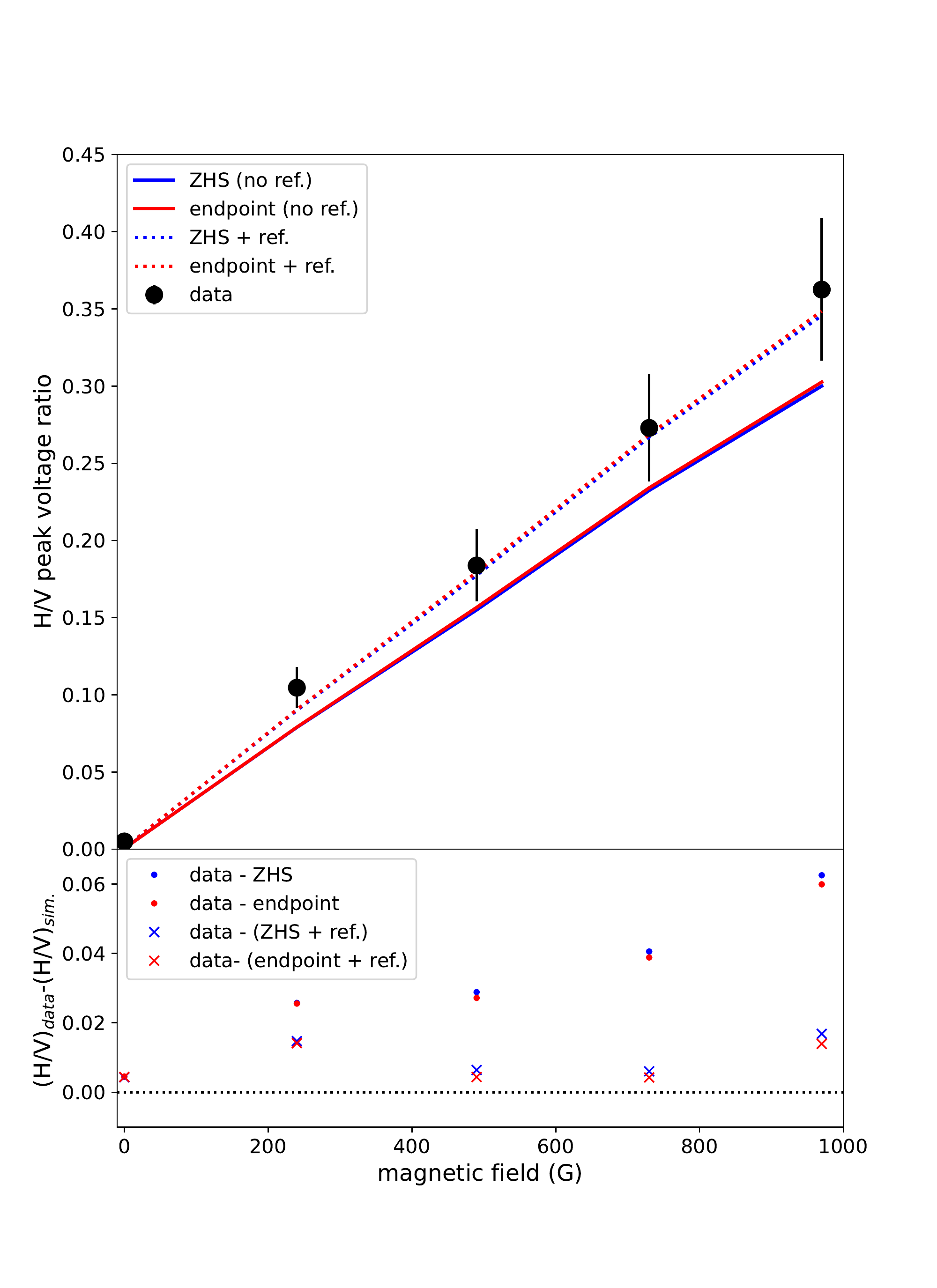}
}
\caption{Top: The ratio of horizontally to vertically polarized signal is shown for an antenna 652~cm above beam height. Data is represented with black points.  The error bars represent the systematic uncertainty on the measurement. Simulations with and without reflections added are also shown, with the solid line indicating no reflections, and the simulations with reflections added are show as dashed lines. Bottom: The distribution of residuals  (data - simulation) is shown.  The solid points in blue and red represent (data - simulation without reflections added) for ZHS and endpoint formalisms respectively, and the crosses represent (data - simulation with reflections added) for ZHS and endpoint formalisms.}\label{fig:Blinearity}
\end{figure}

In Fig.~\ref{fig:Blinearity} the ratio of the peak amplitudes of the horizontally polarized component and the vertically polarized component for different magnetic field strengths is shown.  The simulated ratio before the inclusion of reflections is shown as blue and red solid lines, for ZHS and endpoint formalisms respectively.  The simulated ratio was consistently lower than measured data.  The simulated ratio with the inclusion of reflections is shown as blue and red dashed lines, and is in good agreement with the data. It rises linearly with increasing magnetic field strength, demonstrating the linear scaling of magnetically-induced emission with the magnetic field strength.


\subsection{A scan of the Cherenkov cone}

\begin{figure*}
\centering
\vbox{
\includegraphics[width=0.8\textwidth]{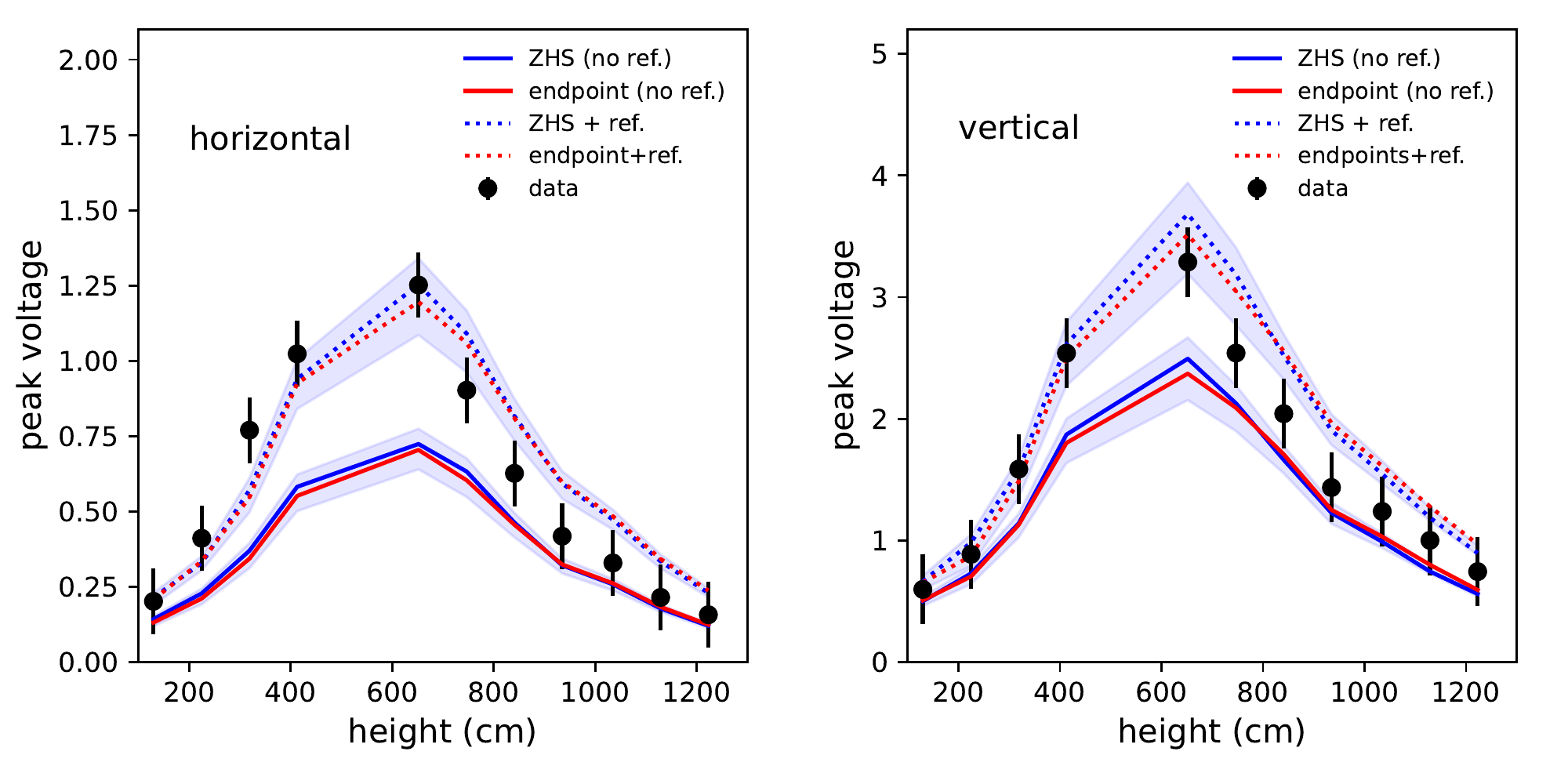}
}
\caption{Maximum signal strength of the horizontally (left) and vertically (right) polarized signal along the vertical axis with a $970\,\mbox{G}$ magnetic field. Data is shown as black dots and the error bars indicate systematic uncertainties.  Simulations with reflections included are indicated with the dashed lines, and without reflections by the solid lines. ZHS simulations are shown in blue and endpoint in red. The light blue shaded region represents the envelope of systematic uncertainties of both endpoint and ZHS simulations.}
\label{fig:Cone_H}
\end{figure*}
The radio emission from a particle shower forms a cone~\cite{Huege:2016veh}. A scan across the Cherenkov cone to measure the angular radiation pattern of Askaryan emission was already performed in a former experiment~\cite{PhysRevLett.99.171101} showing that radiation by a particle shower in a dense medium due to the Askaryan effect forms a cone with its peak at the position determined by the refractive index of the medium.

To measure the angular radiation pattern in the SLAC T-510 experiment, we performed a scan along the vertical axis by placing the antenna tower at different heights.  The agreement of the relative shapes of the measured and simulated cones were presented in~\cite{Belov2016}, however, the absolute signal strength did not agree, due to the fact that reflections were not taken into account in the simulations.  We now include the reflections, and the resulting Cherenkov cone in the $300-900$~MHz band is shown in Fig.~\ref{fig:Cone_H}. We compare the peak amplitude for the vertically (right) as well as the horizontally (left) polarized components of the radio signal along the vertical axis for the measured data and the simulations.  The error bars on the data represent systematic uncertainties on the measurement.  Simulations including reflections are shown as dashed lines, and without reflections are shown as solid lines.  The blue shaded region represents the envelope of systematic uncertainties for both endpoint and ZHS simulations.  Adding reflections improves the agreement between data and simulations overall, and most notably on the Cherenkov cone.

This agreement of the simulations with the data for both polarizations leads to the conclusion that the first-principle simulations can reproduce typical effects which are expected from radio emission from a particle cascade, in particular the Cherenkov-like effects.  The comparison of the data and the simulations shows an accurate prediction of the absolute scale of the radio signal.  Nevertheless, a slight asymmetry of the cones, visible as a shift between the measured cone and the simulated cones, is observable.  This could be explained by diffraction effects in the target, whose impact would change with the height of the antenna, as well as by a more complicated reflection pattern than what was modeled.

\section{Conclusion and Outlook}

The SLAC T-510 experiment is the first experiment that provided a laboratory benchmark for radio frequency emission from electromagnetic showers under the influence of a strong magnetic field.
We compared the measured radio emission from a well-defined particle shower with known primary energy and known beam charge developing in a well-defined target of known geometry to predictions from microscopic simulations which rely on first principles of electrodynamics and have no free parameters.

We chose the parameters of the experiment, such as the target material and the strength of the magnetic field, in such a way that the results of the comparison of data and simulations can be scaled to the relevant frequency ranges for air shower detection~\cite{Belov2016}.
While this experiment does not exactly replicate the physics of an extensive air shower, a good agreement of the predictions and the measurements ensures the applicability of the conclusions to air shower detection.
In addition, since this experiment was a fixed target experiment using a known electromagnetic shower,
it can confirm the validity of the prediction of microscopic calculations with different systematic uncertainties than air shower measurements, in particular without uncertainties in the hadronic interaction models.

Both the endpoint and ZHS formalisms, which we included in the detailed simulation study for the SLAC T-510 experiment, produce the electric field strength for the antenna positions consistent with each other to within $5\%$.  The first comparison of simulation to measured data led to the conclusion that both models can produce the shape of the Cherenkov cone, and that the magnetically-induced radiation scales linearly with magnetic field strength.  However, the agreement of the absolute scales of measurements and simulations disagreed by roughly $35\%$~\cite{Belov2016}.  

In this paper, we demonstrated that an internal reflection at the bottom surface explains the apparent discrepancy between the absolute amplitudes of the measured data and the simulations. We modeled and included this reflection, bringing the difference in the amplitude of Askaryan emission on the Cherenkov cone to within 5\% for the endpoint formalism and 11\% for ZHS.  The agreement for magnetically-induced emission on the cone was even better, at 5\% for the endpoint formalism and less than 1\% for ZHS.  The measurements and simulation now agree within the systematic uncertainties on the Cherenkov cone.  The results of the T-510 experiment demonstrate that microscopic simulations can accurately describe the radio emission from extensive air showers, including their absolute strength, which is a critical result for air shower experiments that detect cosmic rays using radio techniques.

\begin{acknowledgments}
\textbf{\textit{Acknowledgments}} 
The authors thank SLAC National Accelerator Laboratory for providing facilities and support and especially Janice Nelson and Carl Hudspeth for their support and dedication that made T-510 possible. We thank D. Z. Besson for helpful discussions and Clancy W. James for helping to investigate the ZHS fallback threshold. This material is based upon work supported by the Department of Energy under Award Numbers DE-AC02-76SF00515, DE-SC0009937, and others. Work supported in part by grants from the National Aeronautics and Space Administration and the Taiwan Ministry of Science and Technology under project number MOST103-2119-M-002-002, among others. Part of this research was funded through the JPL Internal Research and Technology Development program and through the Frost Fund at the California Polytechnic State University in San Luis Obispo, CA. This work was supported in part by the Kavli Institute for Cosmological Physics at the University of Chicago through grant NSF PHY-1125897 and an endowment from the Kavli Foundation and its founder Fred Kavli. K. Belov acknowledges support from the Karlsruher Institut f\"{u}r Technologie under a guest fellowship. B. Rauch  was supported in part by the McDonnell Center for the Space Sciences at Washington University in St. LouisWe are grateful to the ANITA collaboration for use of antennas and other equipment.
\end{acknowledgments}

\bibliography{SimT510_PRD.bib}

\end{document}